\documentclass[11pt,prd,aps,nofootinbib,preprintnumbers,tightenlines,superscriptaddress]{revtex4}

\usepackage{epsf,graphicx}
\usepackage{times,latexsym,euscript,amstext,amssymb,amsbsy,amsmath,amsfonts}
\usepackage{epsfig}

\usepackage[T1]{fontenc}
\usepackage{charter}
\usepackage{wrapfig}
\usepackage[a4paper,portrait,top=1.5cm, bottom=2.2cm,left=1.5cm,right=1.5cm]{geometry}
\usepackage[colorlinks,citecolor=blue,linktoc=all,linkcolor=cyan]{hyperref}
\usepackage{upgreek}

\def\qt0{\tilde{q}_0}

\def\bq{\boldsymbol{q}}
\def\b0{\boldsymbol{0}}

\def\vare{\varepsilon}

\newcommand{\be}{\begin{equation}}
\newcommand{\ee}{\end{equation}}
\newcommand{\bea}{\begin{eqnarray}}
\newcommand{\eea}{\end{eqnarray}}
\newcommand{\nn}{\nonumber}
\def\eqlab#1{\label{eq:#1}}

\def\seclab#1{\label{sec:#1}}
\def\barr{\left(\begin{array}{c}}
\def\earr{\end{array}\right)}
\def\bmat{\left(\begin{array}{cc}}
\def\emat{\end{array}\right)}
\def\eref#1{(\ref{eq:#1})}
\def\Eqref#1{Eq.~(\ref{eq:#1})}

\def\Figref#1{Fig.~\ref{fig:#1}}

\def\secref#1{Section~\ref{sec:#1}}
\def\hpm{\hphantom{-}}
\DeclareMathOperator{\re}{Re}

\def\dd{{\rm d}}
\def\al{\alpha}

\begin{document}
\preprint{MITP/17-099}

\title{Sum rules across the unpolarized Compton
processes involving generalized polarizabilities and moments of nucleon structure functions}

\author{Vadim Lensky}
\affiliation{Institut f\"ur Kernphysik \& Cluster of Excellence PRISMA, Johannes Gutenberg Universit\"at Mainz, D-55099 Mainz, Germany}
\affiliation{Institute for Theoretical and Experimental Physics, Bol'shaya Cheremushkinskaya 25, 117218 Moscow, Russia}
\affiliation{National Research Nuclear University MEPhI (Moscow Engineering Physics Institute), 115409 Moscow, Russia}

\author{Franziska Hagelstein}
\affiliation{Albert Einstein Center for Fundamental Physics, Institute for Theoretical Physics, University of Bern,  CH-3012 Bern, Switzerland}

\author{Vladimir Pascalutsa}
\affiliation{Institut f\"ur Kernphysik \& Cluster of Excellence PRISMA, Johannes Gutenberg Universit\"at Mainz, D-55099 Mainz, Germany}
\author{Marc Vanderhaeghen}
\affiliation{Institut f\"ur Kernphysik \& Cluster of Excellence PRISMA, Johannes Gutenberg Universit\"at Mainz, D-55099 Mainz, Germany}
\affiliation{Helmholtz Institut Mainz,  D-55099 Mainz, Germany}

\date{\today}

\begin{abstract}
We derive two new sum rules for the unpolarized doubly virtual Compton scattering process on a nucleon, which establish novel low-$Q^2$ relations involving the nucleon's generalized polarizabilities and moments of the nucleon's unpolarized structure functions $F_1(x,Q^2)$ and $F_2(x,Q^2)$. These relations  facilitate the determination of some structure constants which can only be accessed in off-forward doubly virtual Compton scattering, not experimentally accessible at present. 
We perform an empirical determination for the proton and 
compare our results with a next-to-leading-order chiral perturbation
theory prediction. We also show how these relations may be useful for a 
model-independent determination of the low-$Q^2$ 
subtraction function in the Compton amplitude, which enters the 
two-photon-exchange contribution to the Lamb shift of (muonic) hydrogen.
An explicit calculation of the $\Delta(1232)$-resonance contribution
to the muonic-hydrogen $2P-2S$ Lamb shift yields $-1 \pm 1$ $\upmu$eV, confirming the previously conjectured smallness of this effect.
\end{abstract}

\maketitle

\hfill
\tableofcontents

\section{Introduction}

Besides the charge and magnetization distributions in a nucleon, accessed in the elastic lepton-nucleon scattering process, the low-energy nucleon structure is furthermore characterized by its polarizability distributions, which are accessed in Compton scattering (CS) processes with real and virtual photons; see Refs.~\cite{Guichon:1998xv,Drechsel:2002ar,Schumacher:2005an,Holstein:2013kia,Hagelstein:2015egb}
for some reviews. 

The CS process is the starting point 
for deriving sum rules for various electromagnetic
structure quantities~\cite{GellMann:1954db}. For example, the Baldin sum rule for the sum of the dipole polarizabilities~\cite{Baldin}
and the Gerasimov-Drell-Hearn (GDH) 
sum rule for the anomalous magnetic moment~\cite{Gerasimov:1965et,Drell:1966jv}
are derived by considering the real Compton scattering (RCS) process. These sum rules all relate a measured low-energy observable to an integral over a photoabsorption cross section on the nucleon and are thus model-independent relations.  The Burkhardt-Cottingham sum rule~\cite{Burkhardt:1970ti} has been derived for the forward doubly virtual Compton scattering
(VVCS) process, implying that the sum of elastic and inelastic parts of the nucleon's spin structure function $g_2$ integrate to zero for arbitrary photon virtualities. Further sum rules involving 
the spin structure functions were derived by Schwinger \cite{Schwinger:1975ti}.
Another important relation by Cottingham~\cite{Cottingham:1963zz}
connects the unpolarized VVCS with the electromagnetic correction to
the proton-neutron mass difference. It can be used to evaluate
the electromagnetic part of the proton-neutron mass difference~\cite{Gasser:1974wd,Gasser:2015dwa}. Moreover, further assumption that the high-energy behavior of the VVCS amplitude can be parametrized in terms of Reggeon exchanges leads to separate sum rules for each of the two dipole polarizabilities individually~\cite{Gasser:2015dwa}.

Recently, we have presented two sum rules \cite{Pascalutsa:2014zna,Lensky:2017dlc}
which extend the GDH, Burkhardt-Cottingham, and Schwinger family of sum rules. These new sum rules allow us to connect the moments of the nucleon's low-$Q^2$ spin-dependent structure functions $g_{1,2}$, respectively, as measured in inclusive electron scattering~\cite{Kuhn:2008sy,Chen:2010qc}, to  
low-energy electromagnetic structure quantities of the nucleon. The latter can be independently obtained in different experiments: the nucleon's Pauli radius, two of its four lowest-order spin polarizabilities accessed in RCS~\cite{Martel:2014pba}, and the slopes of two of its four lowest-order generalized polarizabilities (GPs), accessed in the virtual Compton scattering (VCS) process~\cite{d'Hose:2006xz,Bourgeois:2006js,Bourgeois:2011zz,Correa:thesis}. 

In the present work, we extend such sum-rule relations to the {\em spin-independent} VVCS process at low $Q^2$. We shall thus derive two new sum rules relating the low-$Q^2$ slopes of the second (first) moments of the nucleon's unpolarized structure functions $F_1$ ($F_2$) to structure constants such as the low-$Q^2$ slope of the nucleon's electric and magnetic GPs and the quadrupole polarizabilities. We will also show how such relations may be useful for a model-independent determination of the low-$Q^2$ ``subtraction function'' in the forward VVCS amplitude $T_1$, 
which affects prominently the two-photon-exchange (TPE) contribution to the Lamb shift of (muonic) hydrogen. 

The paper is organized as follows. In \secref{Formalism}, the general formalism for the spin-independent VVCS, i.e., CS with virtual photons in both the initial and final states, and arbitrary kinematics, is introduced. We discuss the Born contributions as well as the low-energy expansions of the non-Born amplitudes. For the latter, we deduce the limits of RCS [Appendix \ref{RCS}], VCS (where the incoming  photon is virtual and the outgoing photon is real) [Appendix \ref{VCS}], and {\it forward} VVCS. In \secref{VVCS_Sum_Rules}, two new sum rules connecting RCS, VCS, and (forward) VVCS quantities are derived, which are verified in \secref{Verifications} with a next-to-leading-order baryon chiral perturbation theory (BChPT) calculation of the CS polarizabilities. Furthermore, \secref{VVCS_Sum_Rules} introduces a new analyticity constraint on the second derivative of the $T_1(0,Q^2)$ subtraction function, which is verified and studied in \secref{Subtraction} in view of the proton radius puzzle. Empirical and next-to-leading-order (NLO) BChPT predictions for the low-energy coefficients $b_{3,0}$, $b_{4,1}$, and $b_{19,0}$ are derived based on the newly introduced relations and presented in Sections \ref{sec:Verifications} and \ref{sec:Subtraction}, respectively. In \secref{Subtraction}, the effect of the $\Delta(1232)$ excitation on the Lamb shift in muonic hydrogen ($\mu$H) from TPE is evaluated. The paper finishes with a summary and conclusions [\secref{Conclusions}].

\section{Doubly virtual Compton scattering: Spin-independent amplitude}\seclab{Formalism}

\begin{figure}[tb]
\includegraphics[width=0.35\textwidth]{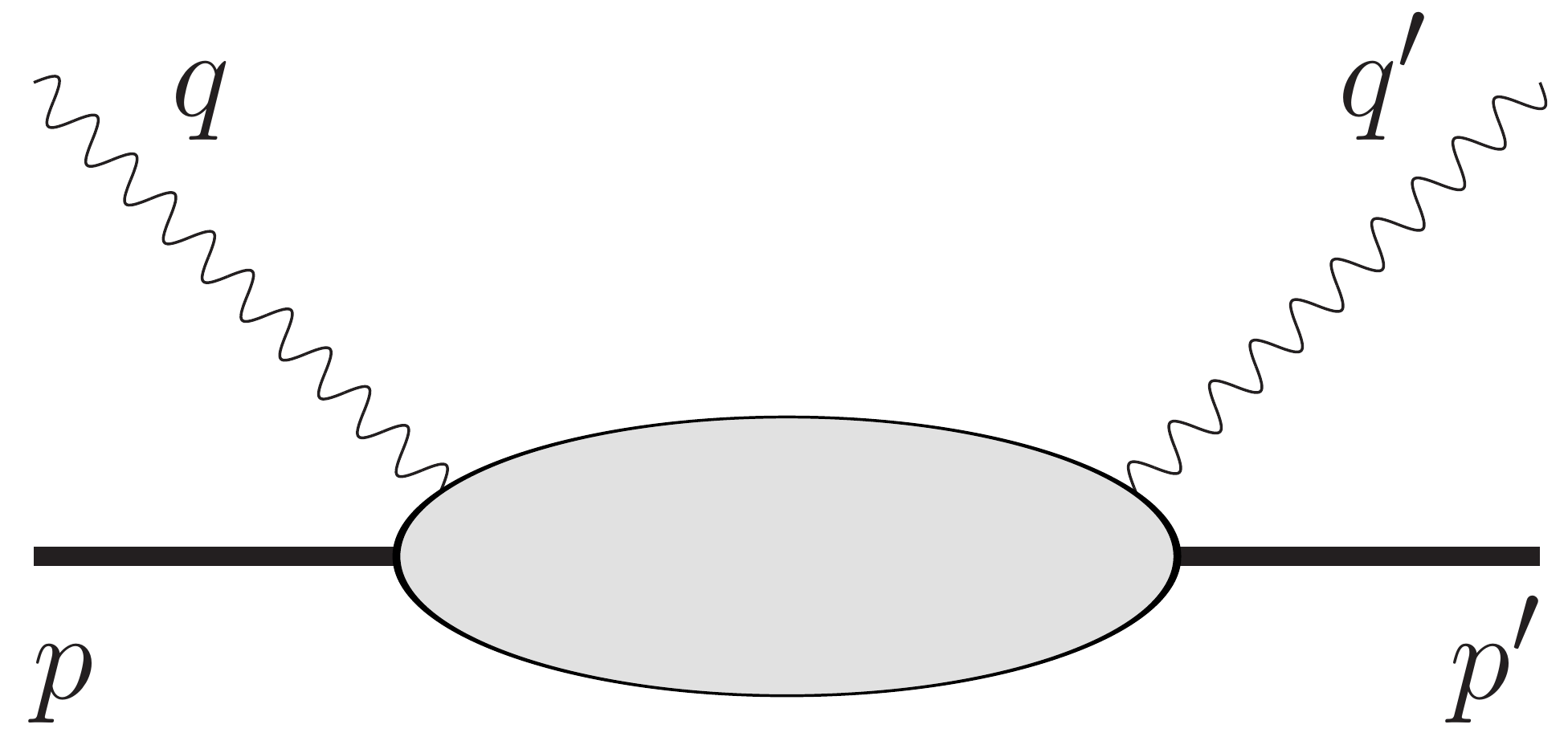}
\caption{Diagram of the Compton scattering process showing the four-momenta of the particles.
}
\label{fig:dvcs}
\end{figure}

The main subject of this work is the Compton scattering
process shown in Fig.~\ref{fig:dvcs}, where the photons are, 
in general, virtual. The spin of the target particle will not play much role in what follows since we will be focusing on the spin-independent observables. Nonetheless, the way the static polarizabilities are defined
(in the rest frame of the target), the recoil corrections may bring
the dependence on the spin polarizabilities, and hence for those effects,
the spin needs to be specified.
Keeping in mind the possible applications
of this formalism, we take the nucleon as the target particle and
hence limit ourselves to the spin-$1/2$ case. 
We therefore consider
the doubly virtual Compton scattering process on the nucleon,
\bea
\gamma^\ast (q, \lambda) + N(p, s) \to \gamma^\ast(q^\prime, \lambda^\prime) + N(p^\prime, s^\prime),
\label{doublevcs}
\eea
where $\lambda$ and $\lambda^\prime $ denote the photon helicities ($0, \pm 1$) 
and  $s$ and $s^\prime$ are the nucleon helicities ($\pm 1/2$).

\subsection{Tensor decomposition}

The VVCS tensor $T^{\mu \nu} $ can be Lorentz decomposed into 18 invariant amplitudes as (in the notation of Ref.~\cite{Drechsel:1997xv})
\bea
M^{\mu \nu} &=& 
\sum_{i \in J} B_i(q^2, q^{\prime \, 2}, q \cdot q^\prime, q \cdot P) \, T_i^{\mu \nu} , 
\quad \quad J = \{ 1,..., 21\} \backslash \{5, 15, 16 \},
\label{eq:dvcs}
\eea
with $P = \frac{1}{2} ( p + p^\prime)$. 
The 18 independent tensors $T_i^{\mu \nu}$ in Eq.~(\ref{eq:dvcs}) are constructed to be gauge invariant~\cite{Tarrach:1975tu}.
Note that in the most general case one has to use the basis consisting of all 21 tensor amplitudes introduced in Ref.~\cite{Drechsel:1997xv} in order to avoid kinematic constraints; however, as long as only the non-Born part of the VVCS amplitude is important (which
is the case in the present work, as detailed below), one can use the 
minimal decomposition of Eq.~\eqref{eq:dvcs}; see Refs.~\cite{Tarrach:1975tu,Drechsel:1997xv}.

The invariant amplitudes $B_i$ depend in general on four kinematic invariants. 
The incoming (outgoing) photon virtualities are denoted by $q^2$ ($q^{\prime \, 2}$), respectively. 
We also define the usual virtualities $Q^2=-q^2$ and $Q^{\prime\, 2}=
-q^{\prime\, 2}$ that are positive for spacelike virtual photons.
These two definitions of a photon's virtuality can be used interchangeably, multiplying the appropriate sign factors where needed.
Furthermore, the variable $q \cdot q^\prime = (q^2 + q^{\prime \, 2} - t)/2$ is related with 
the momentum transfer to the nucleon, $t \equiv (p^\prime - p)^2$. The 
crossing symmetric variable $q \cdot P \equiv M \nu$, 
with $M$ the nucleon mass, can be expressed in  
terms of the Mandelstam variables $s$ and $u$: $M \nu \equiv (s - u)/4$. 

As mentioned, we are only interested in the spin-independent case, which is described by five independent tensors,
\bea
M^{\mu \nu}\big|_{\rm{spin \; indep.}} &=& 
B_1 \, T_1^{\mu \nu}  + B_2 \, T_2^{\mu \nu} + B_3 \, T_3^{\mu \nu}  + B_4 \, T_4^{\mu \nu}  + B_{19} \, T_{19}^{\mu \nu} , 
\label{eq:dvcsunpol}
\eea
where the tensors $T_i^{\mu \nu}$ are symmetric under exchange of the two virtual photons and are given by
\bea
T_1^{\mu \nu} &=& - q \cdot q^\prime g^{\mu\nu}+ q^{\prime \mu} q^{\nu} \, , \nonumber \\
T_2^{\mu \nu} &=&  (2 M \nu)^2 \left( - g^{\mu\nu} +  \frac{q^{\prime \mu} q^{\nu}}{q \cdot q^\prime} \right) 
\,- \, 4 q \cdot q^\prime \left( P^{\mu} - \frac{q \cdot P}{q \cdot q^\prime} q^{\prime \mu} \right)   
\left( P^{\nu} - \frac{q \cdot P}{q \cdot q^\prime} q^{\nu} \right)  \, ,  \nonumber \\
T_3^{\mu \nu} &=& q^2 q^{\prime \, 2} g^{\mu\nu} + q \cdot q^\prime  q^{\mu} q^{\prime \nu}  
- q^2 \, q^{\prime \mu} q^{\prime \nu}  - q^{\prime \, 2} \, q^{\mu} q^{\nu}  \, , \nonumber \\
T_4^{\mu \nu} &=&  (2 M \nu) (q^2 + q^{\prime \, 2}) 
\left( g^{\mu\nu} -  \frac{q^{\prime \mu} q^{\nu}}{q \cdot q^\prime} \right) \nonumber \\ 
&+& \,2 \left( P^{\mu} - \frac{q \cdot P}{q \cdot q^\prime} q^{\prime \mu} \right)   
\left( - q^{\prime 2} q^{\nu} + q \cdot q^\prime  q^{\prime \nu} \right)  
\,+ \,2 \left( - q^2 q^{\prime \mu} +  q \cdot q^\prime  q^{\mu} \right)   
\left( P^{\nu} - \frac{q \cdot P}{q \cdot q^\prime} q^{\nu} \right) \, ,  \nonumber \\
T_{19}^{\mu \nu} &=&   4 q^2  q^{\prime \, 2} 
\left( P^{\mu} - \frac{q \cdot P}{q^2} q^{\mu} \right)   
\left( P^{\nu} - \frac{q \cdot P}{q^{\prime \, 2}} q^{\prime \, \nu} \right)  \, .  
\label{eq:dvcstensors}
\eea

The invariant amplitudes $B_i$ have definite transformation properties with respect to photon crossing, as well as 
nucleon crossing combined with charge conjugation~\cite{Drechsel:1997xv}. 
Using the tensors of Eq.~(\ref{eq:dvcstensors}), 
the photon crossing symmetry of the whole amplitude ($\mu \leftrightarrow \nu$, 
$q \leftrightarrow - q^\prime$) leads to the following relations for the invariant amplitudes:
\bea
B_i(q^2, q^{\prime \, 2}, q \cdot q^\prime, q \cdot P) &=& + B_i(q^{\prime 2}, q^{2}, q \cdot q^\prime, - q \cdot P), \quad \quad (i = 1, 2, 3, 19), \nonumber \\
B_i(q^2, q^{\prime \, 2}, q \cdot q^\prime, q \cdot P) &=& - B_i(q^{\prime 2}, q^{2}, q \cdot q^\prime, - q \cdot P), \quad \quad (i = 4). 
\label{eq:gcrossing}
\eea
Furthermore, 
nucleon crossing  combined with charge conjugation ($P \leftrightarrow - P$) leads to the relations
\bea
B_i(q^2, q^{\prime \, 2}, q \cdot q^\prime, q \cdot P) &=& + B_i(q^{2}, q^{\prime 2}, q \cdot q^\prime, - q \cdot P), \quad \quad (i = 1, 2, 3, 19), \nonumber \\
B_i(q^2, q^{\prime \, 2}, q \cdot q^\prime, q \cdot P) &=& - B_i(q^{2}, q^{\prime 2}, q \cdot q^\prime, - q \cdot P), \quad \quad (i = 4). 
\label{eq:ncrossing}
\eea

\subsection{Born contribution}

An important contribution to the nucleon Compton amplitude at low energies corresponds with 
a nucleon intermediate state in the blob of Fig.~\ref{fig:dvcs},  referred to as the Born term. This contribution is, by definition, not
affected by structure-dependent constants, such as polarizabilities. 
The Born term is defined by using the electromagnetic vertex for the transition $\gamma^* (q) + N(p) \to N(p + q)$ given as
\be
\label{emvertex}
\Gamma^\mu \;=\; F_D(q^2) \, \gamma^\mu \;+\;
F_P(q^2) \, i \sigma^{\mu \nu} \frac{q_\nu}{2 M} \, ,
\ee
with $F_D$ and $F_P$ the Dirac and Pauli form factors of nucleon $N$,
normalized as $F_D(0)=e_N$ and $F_P(0)=\kappa_N$,
where $e_N$ is the charge in units of the proton charge $e$ ($e_N = 0$ for the neutron) and $\kappa_N$ is 
the anomalous magnetic moment in units of the nuclear magneton $e/2M$; $\sigma^{\mu\nu} = (i/2)[\gamma^\mu,\gamma^\nu]$.
With this choice, the Born contribution to the spin-independent VVCS amplitudes is given by 
\bea
B_1^\mathrm{Born}  & = &
   \frac{1}{4 M^3} F_P(q^2) F_P(q^{\prime \, 2}) \nonumber \\
&-& \frac{\nu_B}{2 M^2}  \frac{1}{\nu^2 - \nu_B^2 + i \vare }
\left\{  G_M(q^2) G_M(q^{\prime \, 2}) -  F_D(q^2) F_D(q^{\prime \, 2})
+ \frac{q \cdot q^\prime}{4 M^2}  F_P(q^2) F_P(q^{\prime \, 2})
\right\}  \,, \nonumber  \\
B_2^\mathrm{Born}  & = &
 \frac{1}{4 M^3}  \frac{1}{\nu^2 - \nu_B^2 + i \vare }
\left\{ F_D(q^2) F_D(q^{\prime \, 2})
- \frac{q \cdot q^\prime}{4 M^2}  F_P(q^2) F_P(q^{\prime \, 2})
\right\}  \,,  \nonumber  \\
B_3^\mathrm{Born}  & = & 
B_4^\mathrm{Born}   \, =\,
B_{19}^\mathrm{Born} \, =\, 0 ,
\label{dvcsborn}
\eea
where $\nu_B \equiv - q \cdot q^\prime / (2 M)$, and we introduced the Sachs magnetic form factor, $G_M = F_D + F_P$.

The Born contribution of Eq.~(\ref{dvcsborn}) can be split into pole 
and nonpole contributions.  The  
pole contributions (also called elastic contributions) are singular at $\nu = \nu_B$. The only nonpole piece 
in Eq.~(\ref{dvcsborn}) is obviously the first term, i.e., $B_1^\mathrm{np} = F_P(q^2) F_P(q^{\prime \, 2})/4M^3$. The rest 
of the Born terms are the pole contributions.

\subsection{Low-energy expansions}

The non-Born part of the VVCS amplitudes 
(denoted as $\bar B_i$) can be expanded 
for small values of $q^2, q^{\prime \, 2}, q \cdot q^\prime$, and $\nu$, 
with the expansion coefficients given by polarizabilities. 
We use the low-energy expansions (LEXs) in $k = \{q, q^\prime\}$ established in Ref.~\cite{Drechsel:1997xv},
\begin{subequations}
\bea
\bar{B}_i &=& b_{i,0} + b_{i,2a} q \cdot q^\prime + b_{i,2b} ( q^2 + q^{\prime \, 2}) + b_{i,2c} ( 2 M \nu)^2  +  {\cal  O}(k^4), \quad \quad (i = 1, 2, 3, 19), 
\label{lexdvcsa} \\
\bar{B}_i  &=& 2 M \nu \left\{ b_{i,1} + b_{i,3a} q \cdot q^\prime + b_{i,3b} ( q^2 + q^{\prime \, 2}) + b_{i,3c} ( 2 M \nu)^2    + {\cal  O}(k^4) \right\} , \quad \quad (i = 4), 
\label{lexdvcsb}
\eea
\end{subequations}
where the parameters $b_{i, x}$ are structure constants.
We notice that in order to fully specify the low-energy structure of the 
spin-independent doubly virtual Compton amplitude one requires two constants at the lowest order ($b_{1,0}$ and $b_{2,0}$) and nine additional constants when going to the next order: six coefficients arising from higher-order terms in $\bar B_1$ and $\bar B_2$ and the three lowest-order coefficients in the amplitudes $\bar B_3, \bar B_4$, and $\bar B_{19}$, which are the amplitudes which are accompanied by tensor structures of higher order in $k = \{q, q^\prime\}$. 

The RCS process, corresponding with 
$q^2 = q^{\prime \, 2} = 0$, allows one to constrain the two lowest-order coefficients in $\bar B_1$ and $\bar B_2$ as well as four of the next-order coefficients in $\bar B_1$ and $\bar B_2$. We detail the connection between these coefficients and the polarizabilities,  accessible through RCS, in Appendix~\ref{RCS}. These relations are given by (with the
fine-structure constant $\alpha_\mathrm{em}\equiv e^2/4\pi\simeq 1/137$):
\begin{subequations}
\bea
b_{1, 0} &=& \frac{1}{\alpha_\mathrm{em}} \beta_{M1} , 
\label{b10} \\
b_{1, 2a} &=&- \frac{1}{\alpha_\mathrm{em}} \frac{1}{6} \,\beta_{M2} , 
\label{b12a} \\
b_{1, 2c} &=& \frac{1}{\alpha_\mathrm{em}} \frac{1}{(2 M)^2} 
\left[ \beta_{M1, \nu} + \frac{1}{12} (2 \beta_{M2} - \alpha_{E2})  
+ \frac{1}{M} (\gamma_{M1M1} + \gamma_{E1M2})  
\right] ,
\label{b12c}  \\
b_{2, 0} &=& - \frac{1}{\alpha_\mathrm{em}} \frac{1}{(2 M)^2}\, (\alpha_{E1} + \beta_{M1}) , 
\label{b20}  \\
b_{2, 2a} &=& \frac{1}{\alpha_\mathrm{em}} \frac{1}{(2 M)^2} 
\left[ \frac{1}{6} (\alpha_{E2} + \beta_{M2}) 
- \frac{1}{M} (\gamma_{E1M2} + \gamma_{M1E2})  
+ \frac{1}{2 M^2} (\alpha_{E1} + \beta_{M1}) 
\right] , 
\label{b22a}  \\
b_{2, 2c} &=& - \frac{1}{\alpha_\mathrm{em}} \frac{1}{(2 M)^4} 
\left[ \alpha_{E1, \nu} + \beta_{M1, \nu} + \frac{1}{12} (\alpha_{E2} + \beta_{M2})  \right] .
\label{b22c}  
\eea
\end{subequations}
Besides the electric (magnetic) dipole polarizabilities $\alpha_{E1}$ ($\beta_{M1}$),  the above relations involve 
the corresponding electric (magnetic) dispersive polarizabilities $\alpha_{E1, \nu}$ ($\beta_{M1, \nu}$) and  
the electric (magnetic) quadrupole polarizabilities $\alpha_{E2}$ ($\beta_{M2}$). Furthermore, there are 
recoil terms (proportional to $1/M$ relative to the quadrupole polarizability terms), which involve the 
lowest-order nucleon spin polarizabilities $\gamma_{M1M1}$, $\gamma_{E1M2}$, and $\gamma_{M1E2}$, as well as recoil terms (proportional to $1/M^2$), which involve the 
scalar polarizabilities $\alpha_{E1}$ and $\beta_{M1}$.

In Appendix~\ref{VCS}, we show that the nonforward VCS process, corresponding with an outgoing real photon, i.e., $q^{\prime \, 2} = 0$, and an initial spacelike virtual photon with virtuality $q^2$, provides a second limit for the doubly virtual Compton scattering. Its measurement allows us to constrain two more of the next-order coefficients in $\bar B_1$ and $\bar B_2$ as
\begin{subequations}
\bea
b_{1, 2b} &=& 
-\frac{1}{\alpha_\mathrm{em}}\left\{\beta_{M1}^\prime
+\frac{1}{8M^2}\beta_{M1}
\right\}, 
\label{b12b} \\ 
b_{2, 2b} &=&\frac{1}{\alpha_\mathrm{em}}
\frac{1}{(2 M)^2} \left\{
\alpha_{E1}^\prime+\beta_{M1}^\prime 
-  \frac{1}{2 M } 
\left(  \delta_{LT} +  \gamma_{E1E1} - \gamma_{E1M2} \right)
 - \frac{1}{8 M^2} 
 \left( \alpha_{E1} + \beta_{M1} \right) 
\right\},\quad
\label{b22b} 
\eea
\end{subequations}
which involve the slopes at $Q^2 =0$ of the
magnetic ($\beta_{M1}^\prime$) and electric ($\alpha_{E1}^\prime$) GPs, defined through Eqs.~(\ref{GPslopeM}) and (\ref{GPslopeL}). Furthermore, the recoil terms (proportional to $1/M$ and $1/M^2$ relative to 
$\beta_{M1}^\prime$ or $\alpha_{E1}^\prime$) 
involve, besides $\alpha_{E1}$ and $\beta_{M1}$, also the RCS spin polarizabilities $\gamma_{E1E1}$, $\gamma_{E1M2}$, as well as the longitudinal-transverse spin polarizability $\delta_{LT}$ at $Q^2 = 0$, which is accessed from a moment of the nucleon spin-dependent structure functions $g_1$ and $g_2$. 

We note that all quantities entering the {\it rhs} of Eqs.~(\ref{b10})-(\ref{b22c}) and 
Eqs.~(\ref{b12b})-(\ref{b22b}) 
are observables which are accessed either through the RCS process, VCS process, or forward structure functions. 

\subsection{Forward limit}\seclab{fVVCS}

Besides the low-energy RCS and VCS processes, we can consider as another limit of the doubly virtual Compton process of Eq.~(\ref{doublevcs})  
the {\it forward} VVCS limit, 
which corresponds with $q^\prime = q$ and $p^\prime = p$. Notice that for this process 
$q^2 = q^{\prime \, 2} = q \cdot q^\prime = - Q^2 < 0$. 
The helicity averaged forward VVCS process is described by two invariant amplitudes, denoted by $T_1$ and $T_2$, which are functions of two kinematic invariants: $Q^2$ and $\nu$. 
Its covariant tensor structure can be written as
\bea
\label{vvcs}
\alpha_\mathrm{em}  \, M^{\mu \nu}(\mathrm{VVCS}) \big|_{\mathrm{spin \, indep.}}&\equiv&  
 \left( g^{\mu\nu}-\frac{q^{\mu}q^{\nu}}{q^2}\right) T_1(\nu, Q^2) 
- \frac{1}{M^2} \left(p^{\mu}-\frac{p\cdot
q}{q^2}\,q^{\mu}\right) \left(p^{\nu}-\frac{p\cdot
q}{q^2}\, q^{\nu} \right) T_2 (\nu, Q^2) , \qquad%\nonumber \\
\eea
where $\alpha_\mathrm{em}$ is conventionally 
introduced in defining the forward amplitudes $T_1$ and $T_2$. 
The optical theorem relates the imaginary parts of $T_1$ and $T_2$ 
to the two unpolarized structure functions of inclusive electron-nucleon scattering as
\bea
\label{optical}
{\rm{Im}}\ T_1(\nu,\,Q^2) = \frac{e^2}{4 M} F_1(x,\,Q^2) \, , \quad
{\rm{Im}}\ T_2(\nu,\,Q^2)  = \frac{e^2}{4 \nu}  F_2(x,\,Q^2) \, ,  
\eea
where $x \equiv Q^2 / 2 M \nu$ and where 
$F_1$ and $F_2$ are the conventionally defined structure functions parametrizing 
inclusive electron-nucleon scattering. 
The imaginary parts of the forward scattering amplitudes, Eqs.~(\ref{optical}),
get contributions from both elastic scattering at $\nu = \nu_B  \equiv Q^2/(2M)$, or equivalently $x=1$, as well as 
from inelastic processes above the pion production threshold, corresponding with $\nu > \nu_0 \equiv m_\pi + (Q^2 + m_\pi^2)/(2 M)$ with $m_\pi$ the pion mass,
or equivalently $x < x_0 \equiv Q^2 / (2 M \nu_0)$. 

Expressing the doubly virtual Compton tensors of Eq.~(\ref{eq:dvcstensors}) in the forward limit, 
the VVCS amplitudes $T_1$ and $T_2$ can be readily expressed in terms of the $B_i$ amplitudes of  Eqs.~(\ref{eq:dvcsunpol}) as
\bea
T_1(\nu, Q^2) &=& \alpha_\mathrm{em} \, \left\{ Q^2 \, B_1 - 4 M^2 \nu^2 \, B_2 + Q^4 \, B_3 - 4 M \nu Q^2 \, B_4 \right\}, 
\label{rel1} \\
T_2(\nu, Q^2) &=& \alpha_\mathrm{em} \, 4 M^2 Q^2 \left\{ - B_2 - Q^2 B_{19} \right\}, 
\label{rel2} 
\eea
where the amplitudes $B_i$ also depend on $\nu$ and $Q^2$ for forward kinematics. 

Using Eq.~(\ref{dvcsborn}), we can express the Born contributions in the forward limit as
\bea
\label{vvcsborn}
T_1^\mathrm{Born}  & = &
 -  \frac{\alpha_\mathrm{em}}{M}
\left( F_D^2(q^2) + \frac{\nu_B^2}{\nu^2-\nu_B^2+i \vare} \,G_M^2(q^2) \right) \,, \nonumber  \\
T_2^\mathrm{Born} & = &
- \frac{\alpha_\mathrm{em}}{M} \,\frac{Q^2}{\nu^2-\nu_B^2+i \vare} \,
\left( F_D^2(q^2) + \frac{Q^2}{4 M^2} \, F_P^2(q^2) \right) \, , 
\eea
and the corresponding pole parts as
\bea
\label{vvcspole}
{\rm{Re}}\ T_1^\mathrm{pole}  & = &
 -  \frac{\alpha_\mathrm{em}}{M}
 \frac{\nu_B^2}{\nu^2-\nu_B^2} \,G_M^2(q^2)  \,, \nonumber  \\
{\rm{Re}}\ T_2^\mathrm{pole} & = &
- \frac{\alpha_\mathrm{em}}{M} \,\frac{Q^2}{\nu^2-\nu_B^2} \,
\left( F_D^2(q^2) + \frac{Q^2}{4 M^2} \, F_P^2(q^2) \right) \, . 
\eea

Using the LEXs of the non-Born amplitudes $\bar B_i$, given in 
Eqs.~(\ref{lexdvcsa}) and (\ref{lexdvcsb}), we can obtain 
from Eqs.~(\ref{rel1}) and (\ref{rel2}) LEXs for the non-Born parts $\bar T_{1, 2}$ of the amplitudes $T_{1,2}$. Up to fourth order in $k = \{\nu, Q\}$, 
these LEXs are given by
\bea
\bar{T}_1(\nu, Q^2) &=& \alpha_\mathrm{em} \, \left\{ Q^2 \, b_{1, 0} - 4 M^2 \nu^2 \, b_{2, 0} 
+ Q^4 \left[ - b_{1, 2a} - 2 b_{1, 2b} + b_{3, 0} \right] \right. \nonumber \\
&& \left. \hspace{0.75cm}
- (2 M \nu)^4 b_{2, 2c} + (2 M \nu)^2 Q^2 \left[ b_{1, 2c} + b_{2, 2a} + 2 b_{2, 2b} - 2 b_{4, 1} \right] 
\right\} + {\cal O}(k^6) , 
\label{lexvcs1} \\
\bar{T} _2(\nu, Q^2) &=& - \alpha_\mathrm{em} \,4 M^2 Q^2 \,  
\left\{  b_{2, 0} + Q^2 \left[ - b_{2, 2a} - 2 b_{2, 2b} + b_{19, 0} \right] 
+ (2 M \nu)^2  b_{2, 2c} \right\}
+  {\cal O}(k^6) . 
\label{lexvcs2}
\eea
Besides the low-energy coefficients constrained from RCS and VCS, as given in Eqs.~(\ref{b10}) and (\ref{b22c}) and 
Eqs.~(\ref{b12b}) and (\ref{b22b}), the knowledge of the amplitudes 
$\bar T_1$ and $\bar T_2$ to fourth order requires in addition the knowledge of the constants $b_{3,0}$, $b_{4,1}$, and $b_{19,0}$, which we will discuss in the next section.

In the following, we will also be interested in the amplitude $\bar{T}_1$ at zero energy ($\nu = 0$), which 
plays the role of a subtraction function in a dispersive framework for the VVCS amplitude. 
From Eq.~(\ref{lexvcs1}), we see that its LEX can be expressed as
\bea
\bar{T}_1(0, Q^2) &=& \alpha_\mathrm{em} \, \left\{ Q^2 \, b_{1, 0} 
+ Q^4 \left[ - b_{1, 2a} - 2 b_{1, 2b} + b_{3, 0} \right] 
\right\} + {\cal O}(Q^6) . 
\label{lexvcs3} 
\eea

\subsection{Sum rules}\seclab{VVCS_Sum_Rules}

Using the RCS constraints of Eqs.~(\ref{b10}) and (\ref{b22c}) and the VCS constraints of Eqs.~(\ref{b12b}) and (\ref{b22b}) on the low-energy coefficients, we can express the spin-independent VVCS amplitudes 
of Eqs.~(\ref{lexvcs1}) and (\ref{lexvcs2}) including all terms up to fourth order in either $Q$ or $\nu$ as
\bea
\bar{T} _1(\nu, Q^2) &=& 
Q^2 \, \beta_{M1} + \nu^2 \, (\alpha_{E1} + \beta_{M1})  
+ \nu^4 \left[  \alpha_{E1, \nu} + \beta_{M1, \nu} + \frac{1}{12} (\alpha_{E2} + \beta_{M2})  
\right] \nonumber \\
&+& Q^2 \nu^2 \left[ \beta_{M1, \nu} + \frac{1}{12} (4 \beta_{M2} + \alpha_{E2}) 
+2 
\left( \alpha_{E1}^\prime +\beta_{M1}^\prime \right)
- 2  \alpha_\mathrm{em} (2M)^2  b_{4, 1} 
   \right. \nonumber \\
&& \left. \hspace{1cm} 
+ \frac{1}{M} \left(- \delta_{LT} + \gamma_{M1M1} - \gamma_{E1E1} - \gamma_{M1E2} +  \gamma_{E1M2} 
\right)
+ \frac{1}{(2 M)^2} (\alpha_{E1} + \beta_{M1}) 
\right] \nonumber \\
&+& Q^4   \, \left[ \frac{1}{6} \beta_{M2} 
+2\beta_{M1}^\prime   
+ \alpha_\mathrm{em} b_{3, 0}
+ \frac{1}{(2 M)^2} \beta_{M1} 
  \right] 
 + {\cal O}(k^6) , 
\label{lexvcs1_exp} \\
\bar{T} _2(\nu, Q^2) &=& Q^2 \, (\alpha_{E1} + \beta_{M1})  
+ Q^2 \nu^2 \left[  \alpha_{E1, \nu} + \beta_{M1, \nu} + \frac{1}{12} (\alpha_{E2} + \beta_{M2})  \right]   
\nonumber \\
&+& Q^4 \left[ \frac{1}{6} (\alpha_{E2} + \beta_{M2}) 
+2 
\left( \alpha_{E1}^\prime +\beta_{M1}^\prime \right)
-  \alpha_\mathrm{em} \,(2 M)^2 b_{19, 0}
\right.  \nonumber \\ 
&& \left. \hspace{0.5cm} 
- \frac{1}{M} \left( \delta_{LT}  + \gamma_{E1E1} + \gamma_{M1E2} \right)  
+ \frac{1}{(2 M)^2} (\alpha_{E1} + \beta_{M1}) 
 \right] 
+  {\cal O}(k^6) . 
\label{lexvcs2_exp}
\eea
We notice that the quadratic terms are fully determined by the proton electric ($\alpha_{E1}$) 
and magnetic ($\beta_{M1}$) dipole polarizabilities. 
The terms of order $\nu^4$ in $\bar T_1$ and of order $ Q^2 \nu^2$ in $\bar T_2$ 
are also fully determined by the electric and magnetic dispersive and quadrupole polarizabilities 
which are observables in RCS.
The term of order $ Q^2 \nu^2$ in $\bar T_1$ involves in addition the slopes at $Q^2 = 0$ of the electric and magnetic GPs, as well as the RCS spin polarizabilities and the longitudinal-transverse spin polarizability $\delta_{LT}$, all of which are also observable quantities either through RCS, VCS, or using moments of spin structure functions. The only unknown in this $Q^2 \nu^2$ term arises from the low-energy coefficient $b_{4,1}$. This term could in principle also be accessed from the VCS process through the LEX of the amplitude $f_3$, as given by Eq.~(\ref{b4vcs}), using the LEX of 
Eq.~(\ref{lexdvcsb}) as
\bea
b_{4, 1} = \frac{1}{2 M} \frac{\rm d\hphantom{\nu}}{{\rm d} \nu} 
 \bar{f}_3(0,0,M \nu)  \bigg|_{\nu =0},
 \label{eq:b41gp}
\eea
by using, e.g., a BChPT calculation for the VCS process~\cite{Lensky:2016nui}. 
However, it will be difficult to extract this constant empirically as it would involve higher-order GPs which have not been quantified so far. In the following, we will show, however, that a forward sum rule will allow us to fix this term. 

Finally, we notice that the quartic terms of order $Q^4$ 
involve the unknown low-energy coefficients 
$b_{3, 0}$ for $\bar T_1$ and 
$b_{19, 0}$ for $\bar T_2$. These coefficients cannot be obtained from RCS or VCS because the corresponding tensors vanish when one or both photons are real. In this section, we will show that 
$b_{19, 0}$ can also be determined from a forward sum rule, involving the longitudinal electroabsorption cross section on a proton. The only unknown parameter which remains then is 
$b_{3, 0}$.  
Its determination will require an observable from the 
doubly virtual Compton process. 
 
Having established the LEXs of the non-Born parts of the forward VVCS amplitudes $T_1$ and $T_2$,
we are ready to use the analyticity in  $\nu$,  for fixed spacelike momentum transfer $q^2 = - Q^2 \leq 0$.
Both amplitudes are even functions of $\nu$. We will present the relations for the nonpole parts of the amplitudes,  $T_1^\mathrm{np}(\nu,\,Q^2)= T_1(\nu,\,Q^2) - T_1^\mathrm{pole}(\nu,\,Q^2)$; i.e., the
well-known pole amplitudes given by Eq.~(\ref{vvcspole}) are subtracted from the full amplitudes.

\subsubsection{Spin-independent amplitude $T_1$}

The dispersion relation (DR) for $T_1$ requires one subtraction, which we take at $\nu = 0$, in order to ensure high-energy convergence, 
\bea
\re T_1^\mathrm{np}(\nu,\,Q^2)\, & = & \,
T_1^\mathrm{np}(0,\,Q^2) + \frac{\nu^2}{2\pi}\,{\mathcal{P}}\,
\int_{\nu_0}^{\infty}\, \mathrm{d}\nu^\prime \, \frac {1}{\nu^\prime (\nu^{\prime \,2} -\nu^2)}\, \frac{e^2}{M} F_1(x^\prime, Q^2),
\label{eq:T1dr} 
\eea
with $x^\prime \equiv Q^2 / (2 M \nu^\prime)$. 
Because the nonpole amplitudes are analytic
functions of $\nu$, they can be expanded in a Taylor series around  $\nu=0$ with
a convergence radius determined by the lowest singularity, the threshold of pion production
at $\nu=\nu_0$. 
Analogous to the low-energy expansion of RCS,
the series in $\nu$, at fixed value of $Q^2$, for  forward VVCS takes the following form~\cite{Drechsel:2002ar},
\bea
\re  T_1^\mathrm{np}(\nu,\,Q^2) \, & = &  T_1^\mathrm{np}(0,\,Q^2)
\,+\, M^{(2)}_{1}(Q^2)  \, \nu^2 \,+\, M^{(4)}_{1}(Q^2)  \, \nu^4 \,+\,{\mathcal{O}}(\nu^6) \,,
\label{eq:T1lex}
\eea
where $M^{(2)}_{1}(Q^2) $ and $M^{(4)}_{1}(Q^2) $ can, respectively, be expressed through the second and fourth moments of the unpolarized nucleon structure function $F_1$ as
\bea
M^{(2)}_{1}(Q^2) & = & \frac{e^2 (2 M)}{\pi \, Q^4}\,\int_{0}^{x_0} \mathrm{d}x^\prime \, x^\prime \,F_1(x^\prime,\,Q^2) 
= \frac{1}{2 \pi^2} \int_{\nu_0}^{\infty}\, \frac{\mathrm{d}\nu^\prime}{\nu^{\prime \,2}} \, \frac{K}{\nu^\prime} \, \sigma_T(\nu^\prime, Q^2)
\, ,
\label{eq:generalbaldin} \\
M^{(4)}_{1}(Q^2) & = & \frac{e^2 (2 M)^3}{\pi \, Q^8}\,\int_{0}^{x_0} \mathrm{d}x^\prime \, x^{\prime \, 3} \,F_1(x^\prime,\,Q^2)
= \frac{1}{2 \pi^2} \int_{\nu_0}^{\infty}\, \frac{\mathrm{d}\nu^\prime}{\nu^{\prime \,4}} \, \frac{K}{\nu^\prime} \, \sigma_T(\nu^\prime, Q^2)
\, .
\label{eq:generalbaldin2} 
\eea
Furthermore, in the second equalities of Eqs.~(\ref{eq:generalbaldin}) and (\ref{eq:generalbaldin2}), we introduced the transverse electroabsorption cross section ($\sigma_T$) on a nucleon through
\bea
K  \sigma_T(\nu^\prime, Q^2 ) 
= \frac{e^2 \pi}{M}  F_1(x^\prime, Q^2),
\eea 
where $K$ is a conveniently defined virtual photon flux factor; e.g., in the definition 
by Hand~\cite{Hand:1963bb}, it is given by $K = \nu^\prime ( 1 - x^\prime)$. 

To obtain the low-energy expansion of the nonpole part $T_1^\mathrm{np}$ 
entering Eq.~(\ref{eq:T1dr}), 
we also need to account for the difference between the Born and pole parts, which can be easily read off 
Eq.~(\ref{vvcsborn}) as
\begin{eqnarray}
T_1^\mathrm{Born}(\nu,Q^2) - T_1^\mathrm{pole}(\nu,Q^2) &=& - \frac{\alpha_\mathrm{em}}{M} F_D^2 
\nonumber \\ 
&=& - \frac{\alpha_\mathrm{em} }{M}  + \frac{\alpha_\mathrm{em} }{3 M} \langle r_1^2 \rangle  Q^2 
- \frac{\alpha_\mathrm{em} }{M} \left( \frac{1}{36} \langle r_1^2 \rangle^2 +  F_D^{\prime\prime}(0) 
\right)  Q^4
+ \mathcal{O}(Q^6), \quad  
\label{eq:T1bmp}
\end{eqnarray}
where $\langle r_1^2 \rangle$ is the squared Dirac radius of the proton and where the $Q^4$ term 
involves the curvature of the Dirac form factor at $Q^2 = 0$, defined as
\bea
F_D^{\prime\prime}(0) \equiv \frac{{\rm d^2} F_D(Q^2 ) }{{\rm d}  (Q^2)^2}   \bigg|_{Q^2 =0} .
\eea
As the difference between the Born and pole term contributions to $T_1$ is independent of $\nu$, it can be fully absorbed in the subtraction function. 
The non-Born part  of the subtraction function $\bar T_1(0,Q^2)$ can be read off Eq.~(\ref{lexvcs1_exp}) as
\bea
\bar{T} _1(0, Q^2) &=& 
\beta_{M1} \, Q^2  
+ \left( \frac{1}{6} \beta_{M2} 
+2\beta_{M1}^\prime  
 + \alpha_\mathrm{em} b_{3, 0}
 + \frac{1}{(2 M)^2} \beta_{M1} 
 \right) \, Q^4   
 + {\cal O}(Q^6) .
 \label{eq:T1nb}
\eea
Apart from  the well-known fact that the expansion of $\bar{T}_1(0,Q^2)$ in 
powers of $Q^2$ starts from the term $\beta_{M1}Q^2$, this relation constrains the next term in the expansion, proportional to $Q^4$:
\bea
\frac{1}{2}\frac{\rm d^2 \bar{T}_1(0,Q^2)}{{\rm d} (Q^2)^2}\bigg|_{Q^2 =0}
\equiv \frac{1}{2}\bar{T}_1^{\prime\prime}(0)
=\frac{1}{6} \beta_{M2} 
+2\beta_{M1}^\prime  
 + \alpha_\mathrm{em} b_{3, 0}
 + \frac{1}{(2 M)^2} \beta_{M1}  \,.
\label{eq:T1pp}
\eea
Combining Eqs.~(\ref{eq:T1bmp}) and (\ref{eq:T1nb}), the subtraction function $T_1^\mathrm{np}(0,Q^2)$ entering the DRs of Eq.~(\ref{eq:T1dr}), including terms up to order $\mathcal{O}(Q^4)$, is given by
 \begin{eqnarray}
T_1^\mathrm{np}(0,Q^2) 
&=&  - \frac{\alpha_\mathrm{em} }{M}  
+ \left[ \beta_M  + \frac{\alpha_\mathrm{em} }{3 M} \langle r_1^2 \rangle \right] Q^2 
\nn \\
&+& \left[ \frac{1}{6} \beta_{M2} 
+2\beta_{M1}^\prime 
+ \alpha_\mathrm{em} b_{3, 0} 
+ \frac{1}{(2 M)^2} \beta_{M1}  
- \frac{\alpha_\mathrm{em} }{M} \left( \frac{1}{36} \langle r_1^2 \rangle^2 +   F_D^{\prime\prime}(0) \right)  \right] Q^4 
+ \mathcal{O}(Q^6). \quad\quad
\label{eq:T1np0}
\end{eqnarray}
In order to completely fix the term of 
$\mathcal{O}(Q^4)$ in the subtraction function, one needs to 
determine the low-energy coefficient $b_{3, 0}$:
\bea 
 b_{3, 0} &=& 
\frac{1}{\alpha_\mathrm{em}  } 
\left\{\frac{1}{2}\bar{T}_1^{\prime\prime}(0)-\frac{1}{6} \beta_{M2} 
-2\beta_{M1}^\prime 
 - \frac{1}{(2 M)^2} \beta_{M1}\right\}.
\label{b30Formula}
\eea
Its determination requires a measurement of the doubly virtual Compton process with a spacelike initial and timelike final photon.  

The $\nu$-dependent terms in the expansion of Eq.~(\ref{eq:T1lex}) can all be determined from sum rules in terms of electroabsorption cross sections on a nucleon, as given, e.g., by 
Eqs.~(\ref{eq:generalbaldin}) and (\ref{eq:generalbaldin2}). 
For $Q^2 = 0$, one can use the LEX of Eq.~(\ref{lexvcs1_exp}) to obtain the Baldin sum rule~\cite{Baldin} and 
a higher-order generalization thereof as
\bea
\alpha_{E1} + \beta_{M1} &=& M^{(2)}_{1}(0) = 
\frac{1}{2 \pi^2} \, \int_{\nu_0}^{\infty}\, \mathrm{d}\nu^\prime \, \frac{\sigma_T(\nu^\prime)}{\nu^{\prime \,2} } , 
\label{eq:baldin} \\
\alpha_{E1, \nu} + \beta_{M1, \nu} + \frac{1}{12} (\alpha_{E2} + \beta_{M2})   &=& M^{(4)}_{1}(0) = 
\frac{1}{2 \pi^2} \, \int_{\nu_0}^{\infty}\, \mathrm{d}\nu^\prime \, \frac{\sigma_T(\nu^\prime)}{\nu^{\prime \,4} } , 
\label{eq:higherbaldin}  
\eea
where $\sigma_T(\nu^\prime)$ is the total photoabsorption cross section on a proton. 

We can next write down a new generalized Baldin 
sum rule for the term proportional to $Q^2 \nu^2$ in the LEX of 
Eq.~(\ref{lexvcs1_exp}):
\bea
\frac{{\rm d}\, M^{(2)}_1(Q^2 )}{{\rm d } Q^2}  \bigg|_{Q^2 =0} \equiv M^{ (2)\prime}_1(0)  &=& 
 \beta_{M1, \nu} + \frac{1}{12} (4 \beta_{M2} + \alpha_{E2}) 
 +2 
\left( \alpha_{E1}^\prime +\beta_{M1}^\prime \right)
- 2  \alpha_\mathrm{em} (2M)^2  b_{4, 1}
  \nonumber \\
&+& \frac{1}{M} \left(- \delta_{LT} + \gamma_{M1M1} - \gamma_{E1E1} - \gamma_{M1E2} +  \gamma_{E1M2} \right) 
+ \frac{1}{(2 M)^2} (\alpha_{E1} + \beta_{M1}). \quad \quad
\label{eq:mp12}
\eea
The structure function moment $M^{(2)}_1(Q^2)$ is an observable and has been measured at the Jefferson Laboratory (JLab)~\cite{Liang:2004tk}. 
If one could determine the low-energy coefficient $b_{4, 1}$ from the VCS process using 
Eq.~(\ref{eq:b41gp}), the sum rule of Eq.~(\ref{eq:mp12}) provides an exact nonperturbative relation 
which relates observables in RCS, VCS, and VVCS. 
A direct determination of $b_{4, 1}$, however, involves higher-order GPs, which may be quite complicated 
to extract from experiment. In practice, one can use the measured value on the {\it lhs} of the sum rule of  Eq.~(\ref{eq:mp12}) in order to determine the low-energy coefficient $b_{4, 1}$ as
\bea 
 b_{4, 1} &=& 
\frac{1}{\alpha_\mathrm{em} 8 M^2 } 
\left\{- M^{(2)\prime}_1(0)   
+ \beta_{M1, \nu} + \frac{1}{12} (4 \beta_{M2} + \alpha_{E2}) 
 +2 
\left( \alpha_{E1}^\prime +\beta_{M1}^\prime \right)
 \right.  \nonumber \\
&& \left. \hspace{2cm} + \frac{1}{M} \left(- \delta_{LT} + \gamma_{M1M1} - \gamma_{E1E1} - \gamma_{M1E2} +  \gamma_{E1M2} \right) 
+ \frac{1}{(2 M)^2} (\alpha_{E1} + \beta_{M1}) \right\}.
\label{b4}
\eea

\subsubsection{Spin-independent amplitude $T_2$}

For the amplitude $T_2$, which is even in $\nu$,  one can write down an unsubtracted DR in $\nu$:
\bea
\re T_2^\mathrm{np}(\nu,\,Q^2)\, =  \,
 \frac{1}{2 \pi} \,{\mathcal{P}}\,
 \int_{\nu_0}^{\infty}\, \mathrm{d}\nu^\prime \,   \frac{1}{\nu^{\prime \, 2} - \nu^2}  \, e^2 F_2(x^\prime ,\,Q^2)\,.
\label{eq:T2dr}
\eea
For the amplitude $T_2$, there is no difference between the Born and pole contributions, as seen from Eq.~(\ref{vvcsborn}). The expansion of the amplitude $T_2^\mathrm{np}$ at small $k = \{\nu, Q\}$ can 
therefore be directly read off Eq.~(\ref{lexvcs2_exp}). 
By evaluating Eq.~(\ref{eq:T2dr}) at $\nu= 0$, taking its derivative with respect to $Q^2$ at $Q^2 = 0$,  
and using the relation
\bea
\left[ \frac{1}{Q^2} F_2 (x, Q^2) \right]_{Q^2 = 0} = \left[ \frac{1}{Q^2} 2 x F_1 (x, Q^2) \right]_{Q^2 = 0} = 
\frac{1}{\pi \, e^2 } \sigma_T,
\eea 
one recovers from the $Q^2$ term in $\bar T_2$ the Baldin sum rule of Eq.~(\ref{eq:baldin}) 
and from the $Q^2 \nu^2$ term in $\bar T_2$ the higher Baldin sum rule of Eq.~(\ref{eq:higherbaldin}).  

Furthermore, Eq.~(\ref{eq:T2dr}) allows us to express $\bar T_2(0,\,Q^2)$ for general $Q^2$ as
\bea 
\bar T_2(0,\,Q^2) = Q^2 M_2^{(1)}(Q^2),
\label{eq:T20}
\eea 
with $M_2^{(1)}(Q^2)$ the first moment of the structure function $F_2$,
\bea
M^{(1)}_2(Q^2)  &=& 
\frac{e^2 (2 M)}{2 \pi \, Q^4}\,\int_{0}^{x_0} \mathrm{d}x^\prime  \,F_2(x^\prime,\,Q^2) 
\nn \\
&=& \frac{1}{2 \pi^2}  \int_{\nu_0}^{\infty}\, 
\frac{\mathrm{d}\nu^\prime}{\nu^{\prime \, 2}}  
\frac{1}{\left( 1 + \frac{Q^2}{\nu^{\prime \, 2}} \right)}
\frac{K}{\nu^\prime} \left[ \frac{}{} \sigma_T(\nu^\prime, Q^2 ) +  \sigma_L(\nu^\prime, Q^2 )\right],   
\label{eq:m21}
\eea 
where the second identity in Eq.~(\ref{eq:m21}) has been obtained 
by expressing $F_2$ through the sum of transverse ($\sigma_T$) and longitudinal ($\sigma_L$) electroabsorption cross sections on a proton as 
\bea
\frac{K}{\nu^\prime} \left[ \frac{}{}  \sigma_T(\nu^\prime, Q^2 ) +  \sigma_L(\nu^\prime, Q^2 )\right] 
= e^2 \pi \left( 1 + \frac{Q^2}{\nu^{\prime \, 2}} \right) \frac{1}{Q^2} F_2(x^\prime, Q^2).
\eea 
We can then express the low-energy expansion of $\bar T_2$ 
including all terms up to fourth order in $k = \{\nu, Q\}$ as\footnote{Note that $M_1^{(2)}(0)=M_2^{(1)}(0)$.}
\bea
\bar T_2(\nu,\,Q^2) \, & = &  
Q^2 \, M^{(2)}_{1}(0)  
+ Q^2 \nu^2 \, M^{(4)}_{1}(0) 
+ Q^4 \, M^{(1)\prime}_2(0) 
+\,{\mathcal{O}}(k^6) \,,
\label{eq:T2lex}
\eea
where $M^{(2)}_{1}(0)$ and $M^{(4)}_{1}(0)$ are given by 
Eqs.~(\ref{eq:baldin}) and (\ref{eq:higherbaldin}) respectively. The term of order $\mathcal{O}(Q^4)$ involves the first derivative at $Q^2 = 0$ of Eq.~(\ref{eq:m21}) and can be obtained through a sum-rule relation from Eq.~(\ref{lexvcs2_exp}) as
\bea 
\frac{{\rm d}\, M^{(1)}_2(Q^2 )}{{\rm d } Q^2}  \bigg|_{Q^2 =0} \equiv 
M^{(1)\prime}_2(0) &=&  \frac{1}{6} (\alpha_{E2} + \beta_{M2}) 
+2 
\left( \alpha_{E1}^\prime +\beta_{M1}^\prime \right)
-  \alpha_\mathrm{em} \,(2 M)^2 b_{19, 0}
\nonumber \\
&&- \frac{1}{M} \left( \delta_{LT}  + \gamma_{E1E1} + \gamma_{M1E2} \right)  
+ \frac{1}{(2 M)^2} (\alpha_{E1} + \beta_{M1}). 
\label{eq:mp21}
\eea 
As the slope $M^{(1)\prime}_2(0)$ is also an observable, the 
knowledge of it therefore allows us to determine the low-energy coefficient 
$b_{19,0}$ as
\bea 
   b_{19, 0} &=& 
  \frac{1}{\alpha_\mathrm{em} \,4 M^2} \left\{ 
- M^{(1)\prime}_2(0) 
+ \frac{1}{6} (\alpha_{E2} + \beta_{M2}) 
+2 
\left( \alpha_{E1}^\prime +\beta_{M1}^\prime \right)\right. 
\nonumber \\
&&\left. \hspace{2cm} - \frac{1}{M} \left( \delta_{LT}  + \gamma_{E1E1} + \gamma_{M1E2} \right)  
+ \frac{1}{(2 M)^2} (\alpha_{E1} + \beta_{M1}) \right\} . 
\label{b19}
\eea 

Let us note that it is also of interest to use the following combination of structure functions,
\bea 
\bar T_L(\nu,Q^2) \equiv  -  \bar T_1(\nu,Q^2)   +  
\frac{\nu^2+Q^2}{Q^2}\bar T_2(\nu,Q^2),
\eea
as its absorptive part can be related to the longitudinal electroabsorption cross section on a nucleon as
\bea 
\mathrm{Im}\, \bar T_L(\nu, Q^2) = \frac{K}{4 \pi} \sigma_L(\nu, Q^2).
\eea 
Its low-energy expansion, obtained by substituting 
$\bar T_1$ and $\bar T_2$ from  
Eqs.~(\ref{lexvcs1_exp}) and (\ref{lexvcs2_exp}) into the 
above definition, goes as
\bea 
\bar T_L(\nu,Q^2) = Q^2  \alpha_{E1} + Q^2\nu^2 \alpha_L  
+Q^4 \alpha^\prime_{E} + O(k^6), 
\label{eq:TL}
\eea
with 
\bea 
\label{alphaL}
\alpha_L &=& M^{(1)\prime}_2(0)  - M^{(2)\prime}_1(0) 
+ M^{(4)}_1(0) \nn  \\
&=& 
\alpha_{E1, \nu} + \frac{1}{12} (2\alpha_{E2} - \beta_{M2}) + \alpha_\mathrm{em} \, 4 M^2 
\big(2 b_{4,1}- b_{19, 0}\big) - \frac{1}{M} \left(  \gamma_{M1M1}  +  \gamma_{E1M2} 
\right) ,
\eea 
where the last line has been obtained by using Eqs.~(\ref{eq:higherbaldin}), (\ref{eq:mp12}), and (\ref{eq:mp21}).
On the other hand, we recognize that  $\alpha_L$
is the value at $Q^2=0$
of the usual $\alpha_L(Q^2)$. It satisfies the sum rule 
in Eq.~(5.36)  of Hagelstein {\it et~al.}~\cite{Hagelstein:2015egb}, which at $Q^2 = 0$ corresponds with the first line in Eq.~(\ref{alphaL}). 
Furthermore, the term of order $\mathcal{O}(Q^4)$ in Eq.~(\ref{eq:TL}) is given by
\bea 
\alpha^\prime_{E} &=& M^{(1)\prime}_2(0)
- \frac{1}{2}\bar{T}_1^{\prime\prime}(0)
\nn \\
 &=& \frac{1}{6} \alpha_{E2} 
 +2\alpha_{E1}^\prime 
 - \alpha_\mathrm{em} \left(4 M^2 b_{19, 0} + b_{3, 0} \right)
 - \frac{1}{M} \left( \delta_{LT}  + \gamma_{E1E1} + \gamma_{M1E2} \right) + \frac{1}{(2 M)^2} \alpha_{E1} . \quad
\eea

To conclude this section, we would like to note that in Refs.~\cite{Gasser:1974wd,Gasser:2015dwa}  a different choice of basis was used for the purpose of evaluating the Cottingham formula for the proton-neutron mass difference. The basis (denoted here by the superscript GL) used in that work is related to ours as
\begin{equation}
T_1^\text{GL}(\nu,Q^2) = -\frac{M}{\alpha_\mathrm{em}Q^2}\left[T_1(\nu,Q^2)-\frac{\nu^2}{Q^2}T_2(\nu,Q^4)\right],\quad T_2^\text{GL}(\nu,Q^2)=\frac{M}{\alpha_\mathrm{em}Q^2}T_2(\nu,Q^4).
\end{equation}
The LEX of the corresponding non-Born amplitudes $\bar{T}^\text{GL}_1(\nu,Q^2)$ and $\bar{T}^\text{GL}_2(\nu,Q^2)$ reads
\bea
\frac{\alpha_\mathrm{em}}{M}\bar{T}_1^\text{GL}(\nu,Q^2) &=& -
\beta_{M1} - \nu^2 \left[ \beta_{M1, \nu} + \frac{1}{12} (2 \beta_{M2} - \alpha_{E2}) 
- \alpha_\mathrm{em}\, 4M^2 \left(2 b_{4, 1} -b_{19,0}\right)
+ \frac{1}{M} \left( \gamma_{M1M1} +  \gamma_{E1M2} 
\right)
\right] \nonumber \\
&-& Q^2   \, \left[ \frac{1}{6} \beta_{M2} 
+2\beta_{M1}^\prime 
+ \alpha_\mathrm{em} b_{3, 0}
+ \frac{1}{(2 M)^2} \beta_{M1} 
  \right] 
 + {\cal O}(k^4) , \\
\frac{\alpha_\mathrm{em}}{M} \bar{T} _2^\text{GL}(\nu, Q^2) &=& (\alpha_{E1} + \beta_{M1})  
+ \nu^2 \left[  \alpha_{E1, \nu} + \beta_{M1, \nu} + \frac{1}{12} (\alpha_{E2} + \beta_{M2})  \right]   
\nonumber \\
&+& Q^2 \left[ \frac{1}{6} (\alpha_{E2} + \beta_{M2}) 
+2 
\left( \alpha_{E1}^\prime +\beta_{M1}^\prime \right)
-  \alpha_\mathrm{em} \,(2 M)^2 b_{19, 0}
\right.  \nonumber \\ 
&& \left. \hspace{0.5cm} 
- \frac{1}{M} \left( \delta_{LT}  + \gamma_{E1E1} + \gamma_{M1E2} \right)  
+ \frac{1}{(2 M)^2} (\alpha_{E1} + \beta_{M1}) 
 \right] 
+  {\cal O}(k^4) . 
 \eea
To obtain the total amplitudes $T_1^\text{GL}$ and $T_2^\text{GL}$, 
one needs to add the Born terms, which read
\bea
\label{vvcsbornGL}
T_1^\mathrm{GL, \, Born}  & = &
 -  \frac{F_P^2(q^2)}{4 M^2} + \frac{1}{\nu^2-\nu_B^2+i \vare} \, \frac{Q^2}{Q^2 + 4 M^2} \, \left(\vphantom{\frac{1}{2}} G_M^2(q^2) - G_E^2(q^2) \right)  \,, \nonumber  \\
T_2^\mathrm{GL, \, Born} & = &
- \frac{1}{\nu^2-\nu_B^2+i \vare} \,
\left( F_D^2(q^2) + \frac{Q^2}{4 M^2} \, F_P^2(q^2) \right) \, .
\eea
The use of amplitudes $T_{1}^\text{GL}$ and $T_2^\text{GL}$ is equivalent to that of $T_1$ and $T_2$ as far as the quartic constraints
derived in this work are concerned. Indeed, the $\nu^2$-dependent term in $T_{1}^\text{GL}$ and the $Q^2$-dependent term in $T_{2}^\text{GL}$ lead to the two new sum rules of Eqs.~(\ref{eq:mp12}) and (\ref{eq:mp21}) respectively.

\section{Sum-rule verifications in baryon chiral perturbation theory and empirical estimates for the low-energy coefficients}\seclab{Verifications}

In this section, we verify the sum rules derived in 
Eqs.~(\ref{eq:mp12}) and (\ref{eq:mp21}).
For this purpose, we will use a covariant next-to-leading-order BChPT calculation of the non-Born part of the CS process. Furthermore, we will provide empirical estimates for the low-energy coefficients entering the sum rules. 

In several previous works, we have provided next-to-leading-order BChPT results for moments of nucleon structure functions~\cite{Lensky:2014dda}, nucleon polarizabilities entering the RCS process~\cite{Lensky:2009uv,Lensky:2015awa}, and 
generalized polarizabilities entering the VCS process~\cite{Lensky:2016nui}. 
Such calculation is fully predictive at orders  
${\mathcal O}(p^3)$ and ${\mathcal O}(p^4/\varDelta)$. 
The ${\mathcal O}(p^3)$ leading-order (LO) contribution to the polarizabilities and the moments of structure functions comes from the pion-nucleon ($\pi N$) loops, and the ${\mathcal O}(p^4/\varDelta)$ NLO contribution comes from the Delta-exchange ($\Delta$) graph and the pion-Delta ($\pi \Delta$) loops. Here, we estimate the $\Delta(1232)$ 
 effects in the so-called $\delta$ counting~\cite{Pascalutsa:2002pi}, with the Delta-nucleon mass difference $\varDelta = M_\Delta - M$ counted as an intermediate scale, $m_\pi\ll\varDelta\ll M$, so that in the $\delta$ counting $m_\pi/\varDelta\sim\varDelta/M\sim \delta$.
 
The quoted field-theory calculations give predictions for all terms entering Eqs.~\eqref{eq:T1pp}, \eqref{eq:mp12}, and \eqref{eq:mp21}, with the exception of $b_{3,0}$ and $b_{19,0}$. 
The covariant BChPT thus allows us to exactly verify the sum rule \eqref{eq:mp12} for $M_1^{(2)\prime}(0)$ by calculating all entries therein separately. The sum rule \eqref{eq:mp21} for $M_2^{(1)\prime}(0)$ and the constraint \eqref{eq:T1pp} for $T_1^{\prime\prime}(0)$, on the other hand, can be used in order to obtain covariant BChPT
predictions for the unknown coefficients $b_{3,0}$ and $b_{19,0}$. A detailed discussion of \Eqref{T1pp} is postponed to \secref{Subtraction}.

The latter two coefficients can be obtained directly from a calculation of the VVCS process in the off-forward regime. We performed such a calculation for the Delta-exchange graph, extending the CS calculation of that graph to the most general VVCS kinematics, obtaining the respective contributions to $b_{3,0}$ and $b_{19,0}$. This allows us
to verify Eqs.~\eqref{eq:T1pp} and \eqref{eq:mp21}, too, albeit only for the Delta-exchange graph contribution at ${\mathcal{O}}(p^4/\varDelta)$.

An additional remark is in order regarding our calculation of the Delta-exchange graph. As explained in Ref.~\cite{Pascalutsa:2005vq}, the magnetic $\gamma N\Delta$ coupling $g_M$
is complemented by the dipole form factor, inferred from vector meson dominance considerations, and needed phenomenologically for a satisfactory description
of electromagnetic nucleon-Delta transitions,
\begin{equation}
g_M\to\frac{g_M}{\left[1+Q^2/\Lambda^2\right]^2}\,,
\label{eq:DipoleGM}
\end{equation}
with the dipole mass $\Lambda^2=0.71~\text{GeV}^2$. The form factor changes the slopes of the VCS GPs and the VVCS structure function moments which enter the sum rules and the analyticity constraint,
specifically, the values of $\beta^\prime_{M1}$, 
$M_1^{(2)\prime}(0)$, $M_2^{(1)\prime}(0)$, and $\bar{T}_1^{\prime\prime}(0)$.\footnote{Note that the quantities which enter the spin-dependent sum rules considered in our previous work~\cite{Pascalutsa:2014zna,Lensky:2017dlc} are not affected by these form factors.} However, the sum rules and the analyticity constraint are not affected. This can be seen explicitly from the expressions for the respective Delta-exchange contributions. In general, we checked that it is possible to add an arbitrary $Q^2$ dependence to the couplings, e.g., by including form factors, without violating the spin-independent sum rules considered herein, or the constraint on the derivative of the subtraction function $\bar{T}_1^{\prime\prime}(0)$, as discussed in \secref{Subtraction}.

\begin{table}[h]
\begin{tabular}{c|c||c|c|c|c|c|c}
\hline\hline
Source & $M_1^{(2)\prime}(0)$ & $-\alpha_\mathrm{em} 8 M^2 b_{4,1}$ & $\beta_{M1,\nu}$ & $(4\beta_{M2}+\alpha_{E2})/12$ &$2 
\left( \alpha_{E1}^\prime +\beta_{M1}^\prime \right)$
& $1/M$ recoil & $1/M^2$ recoil 
\\
\hline
$\pi N$ loops
& $-0.74$     & $\hpm 2.34$ & $1.78$ & $-1.67$ & $-3.53$     & $\hpm 0.29$ & $0.06$  \\
%\hline
$\pi \Delta$ loops
& $-0.20$     & $\hpm 0.40$ & $0.63$ & $-0.62$ & $-0.56$     & $-0.09$    & $0.03$  \\
%\hline
$\Delta$ exchange
& $\hpm 1.00$  & $-1.58$    & $4.72$ & $-1.44$ & $-1.41$  & $\hpm 0.63$ & $0.08$  \\
\hline
Total BChPT
& $\hpm 0.07\pm 0.4$  & $\hpm 1.17\pm 0.6$ & $7.14\pm 2.5$ & $-3.74\pm 1.0$ & $-5.50\pm 1.2$     & $\hpm 0.83\pm 0.3$ & $0.17\pm 0.01$  \\
\hline\hline
Empirical
& $-1.71$  & $1.62$ & $9.37$ & $-5.77$ & $-7.86$     & $\hpm 0.77$ & $0.16$  \\
& BC fit~\cite{Christy:2007ve} & SR extraction 
& DR~\cite{Holstein:1999uu} 
& DR~\cite{Holstein:1999uu} 
& DR~\cite{Pasquini:2001yy, Drechsel:2002ar} 
& DR~\cite{Drechsel:2002ar, Drechsel:2007if}
& SR~\cite{Hagelstein:2015egb}   \\
\hline\hline
\end{tabular}
\caption{Values of the low-energy coefficients entering the sum rule (\ref{eq:mp12}) for $M_1^{(2)\prime}(0)$, 
all in units of $10^{-4}~\text{fm}^5$. 
The first four rows are different contributions in BChPT: all columns are calculated independently in BChPT, verifying the sum rule. 
Errors are estimated as detailed in Ref.~\cite{Lensky:2015awa}.
The last row corresponds with empirical extractions either through the experimental Bosted-Christy (BC) fit~\cite{Christy:2007ve}, dispersive (DR), or sum rule (SR) estimates as described in the text. The value of $b_{4,1}$ in the last row is the sum rule [Eq.~(\ref{b4})] extraction  using the other values in that row as input.}
\label{tab:sr1}
\end{table}

\begin{table}[h]
\begin{tabular}{c|c||c|c|c|c|c}
\hline\hline
Source & $M_2^{(1)\prime}(0)$ & $-\alpha_\mathrm{em} 4 M^2 b_{19,0}$  & $(\alpha_{E2}+\beta_{M2})/6$ &$2 
\left( \alpha_{E1}^\prime +\beta_{M1}^\prime \right)$
& $1/M$ recoil & $1/M^2$ recoil 
\\
\hline
$\pi N$ loops
& $-1.47$ & $0.95$ & $\hpm 0.86$ & $    -3.53$ & $\hpm 0.20$ & $0.06$  \\
%\hline
$\pi \Delta$ loops
& $-0.26$ & $0.23$ & $\hpm 0.08$ & $    -0.56$ & $    -0.05$ & $0.03$  \\
%\hline
$\Delta$ exchange
& $-1.94$ & $0.01$ & $-0.65$ & $-1.41$ & $\hpm 0.03$ & $0.08$  \\
\hline
Total BChPT
& $-3.68\pm 1.1$ & $1.19\pm 0.3$ & $\hpm 0.29\pm 0.3$ & $    -5.50\pm 1.2$ & $\hpm 0.17\pm 0.04$ & $0.17\pm 0.01$  \\
\hline\hline
Empirical
& $-6.63$ & $0.14$ & $\hpm 0.75$ & $    -7.86$ & $\hpm 0.18$ & $0.16$  \\
& BC fit~\cite{Christy:2007ve} & SR extraction  
& DR~\cite{Holstein:1999uu} 
& DR~\cite{Pasquini:2001yy, Drechsel:2002ar} 
& DR~\cite{Drechsel:2002ar, Drechsel:2007if}
& SR~\cite{Hagelstein:2015egb}  \\
\hline\hline
\end{tabular}
\caption{Values of the low-energy coefficients entering the sum rule~(\ref{eq:mp21}) for $M_2^{(1)\prime}(0)$, 
all in units of $10^{-4}~\text{fm}^5$.
The first four rows are different contributions in BChPT: the $\Delta$-exchange contributions serve as a verification of the sum rule.
Errors are estimated as detailed in Ref.~\cite{Lensky:2015awa}.
The last row corresponds with empirical extractions either through the experimental Bosted-Christy (BC)~\cite{Christy:2007ve}, dispersive (DR), or sum rule (SR) estimates as described in the text. 
The value of $b_{19,0}$ in the last row is the sum rule [Eq.~(\ref{b19})] extraction using the other values in that row
as input.  
}
\label{tab:sr2}
\end{table}

We show the BChPT estimates for all terms entering the sum rule~(\ref{eq:mp12}) for $M_1^{(2)\prime}(0)$ in Table \ref{tab:sr1}, and 
the sum rule~(\ref{eq:mp21}) for $M_2^{(1)\prime}(0)$ in Table \ref{tab:sr2}.
The values in both tables include the contribution of the dipole form factor in the Delta-exchange
graph; values without those contributions can be obtained by adding $4\beta_{M1}^\Delta/\Lambda^2=1.57 \times 10^{-4}~\text{fm}^5$ where appropriate. Remember that $b_{4,1}$ was known from BChPT before~\cite{Lensky:2016nui}, while the coefficient $b_{19,0}$ was previously unknown. For the Delta-exchange contribution, $b_{19,0}$ has been calculated directly from the off-forward VVCS process, whereas the $\pi N$- and $\pi \Delta$-loop contributions were deduced from the sum rule and the BChPT predictions for the remaining quantities in \Eqref{mp21}.

\begin{figure}[h]
\begin{center}
\includegraphics[width=0.7\textwidth]{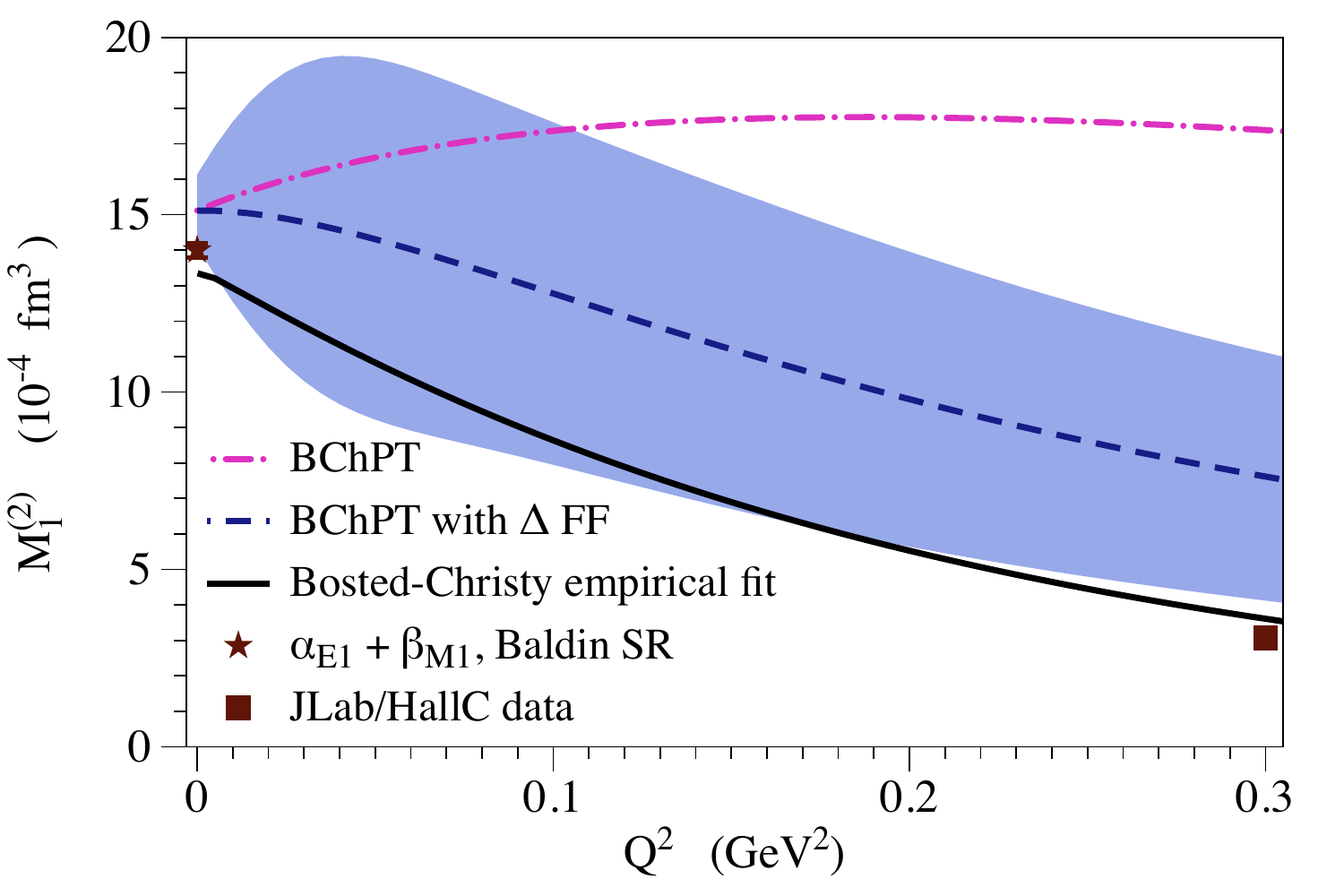}
\caption{$Q^2$ dependence of the proton structure moment 
$M^{(2)}_1$ according to the empirical Bosted-Christy (BC) fit (black solid curve)~\cite{Christy:2007ve}, in comparison with the $\pi N + \Delta + \pi \Delta $ BChPT calculation. For the latter, we also show the result when an additional form factor dependence is included in the $\Delta$-exchange contribution as given by Eq.~\eqref{eq:DipoleGM}; blue dashed (magenta dashed-dotted) curves show the results with (without) the form factor. The blue band shows the uncertainty
of the BChPT result with the form factor, estimated as in Ref.~\cite{Lensky:2016nui}.
At the real photon point, the observable yields the Baldin sum-rule value for $\alpha_{E1} + \beta_{M1}$~\cite{Hagelstein:2015egb}. 
The data point at $Q^2 = 0.3$~GeV$^2$ is from JLab/HallC~\cite{Liang:2004tk}. 
}
\label{fig:momentsf1}
\end{center}
\end{figure}

\begin{figure}[h]
\begin{center}
\includegraphics[width=0.7\textwidth]{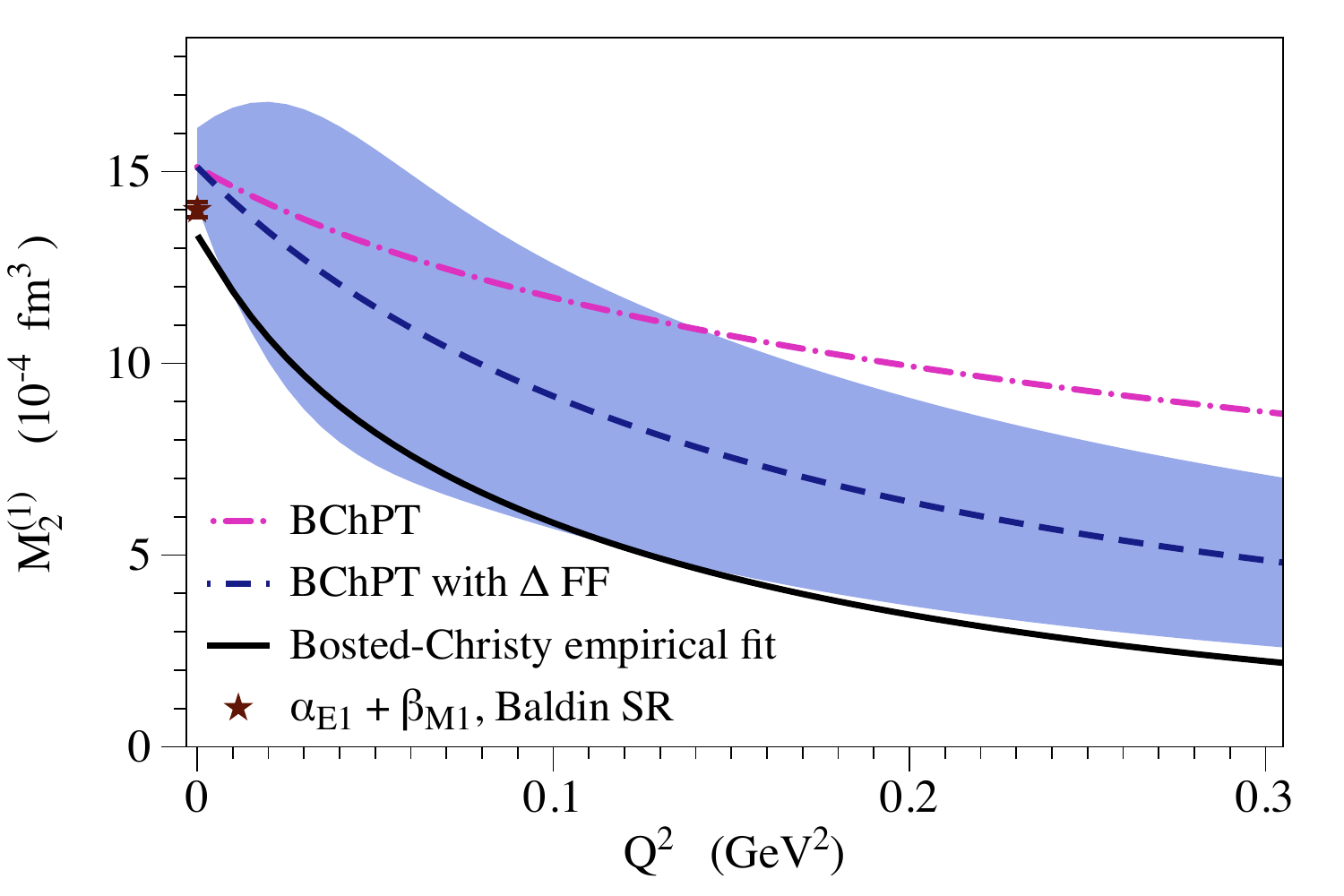}
\caption{$Q^2$ dependence of the proton structure moment 
$M^{(1)}_2$ according to the empirical Bosted-Christy (BC) fit (black solid curve)~\cite{Christy:2007ve}, in comparison with the $\pi N + \Delta + \pi \Delta $ BChPT calculation. For the latter, we also show the result when an additional form factor dependence is included in the $\Delta$-exchange contribution as given by Eq.~\eqref{eq:DipoleGM}; blue dashed (magenta dashed-dotted) curves show the results with (without) the form factor. The blue band shows the uncertainty
of the BChPT result with the form factor, estimated as in Ref.~\cite{Lensky:2016nui}. At the real photon point, the observable yields the Baldin sum-rule value for $\alpha_{E1} + \beta_{M1}$~\cite{Hagelstein:2015egb}. 
}
\label{fig:momentsf2}
\end{center}
\end{figure}

Having verified the sum rules, we can provide empirical estimates of the low-energy coefficients. 
The left-hand sides of both Eq.~(\ref{eq:mp12}) and Eq.~(\ref{eq:mp21}) can be estimated from the measured moments of proton structure functions.  
We show the empirical Bosted-Christy (BC) fit~\cite{Christy:2007ve} for the moments $M_1^{(2)}$ and $M_2^{(1)}$ in the low-$Q^2$ region in Figs.~\ref{fig:momentsf1} and~\ref{fig:momentsf2}, respectively. Their slopes at $Q^2 = 0$ are listed in Tables
\ref{tab:sr1} and \ref{tab:sr2}. 
Furthermore, we use the dispersive estimates of Ref.~\cite{Holstein:1999uu} for the higher-order real Compton polarizabilities and of Refs.~\cite{Pasquini:2001yy, Drechsel:2002ar} 
for the GPs. We   
use the phenomenological MAID2007 fit~\cite{Drechsel:2007if} as input for the $\pi N$-channel contribution in the DRs. 
The recoil terms on the right-hand sides of Eqs.~(\ref{eq:mp12}) and (\ref{eq:mp21}), which are proportional to $1/M$ and $1/M^2$, depend on the lowest-order spin and scalar polarizabilities, respectively. To estimate these terms, we use the empirical values listed in Table~\ref{tab:recoil}. 
Using the sum rules for $M_1^{(2)\prime}(0)$ and 
$M_2^{(1)\prime}(0)$, we are then able to provide empirical estimates for the low-energy coefficients $b_{4,1}$ and $b_{19, 0}$, 
as shown in Tables~\ref{tab:sr1} and \ref{tab:sr2}. 

One can see from Tables~\ref{tab:sr1} and~\ref{tab:sr2} that 
there is a reasonable agreement between the BChPT values and the empirical ones for most terms entering Eqs.~\eqref{eq:mp12}
and~\eqref{eq:mp21}. We see differences for $M_1^{(2)\prime}(0)$, which
is close to zero in BChPT but is negative in the empirical fit, as well as for $b_{19,0}$ which is very small in the empirical extraction. As both of these quantities yield relatively small contributions to the respective sum rules shown in Tables~\ref{tab:sr1} and~\ref{tab:sr2}, the differences can partly be attributed to cancellations between different terms in these relations. Figures~\ref{fig:momentsf1} and~\ref{fig:momentsf2} also demonstrate that there is qualitative
agreement between the BChPT and the BC fit results for $M_1^{(2)}(Q^2)$ and $M_2^{(1)}(Q^2)$--- most of the difference is just due to the different static values of $\alpha_{E1}+\beta_{M1}$ in the two calculations.
Apart from that, the BChPT curves agree, within their (rather wide) error
bands, with the BC fit results.
 
\begin{table}[h]
{\centering 
\begin{tabular}{c|c|c}
\hline
 & Value & Source \\
\hline
\hline
$\alpha_{E1} + \beta_{M1}$  
& \quad $14.0 \pm 0.2$  ($10^{-4}$ fm$^3$) \quad
& Baldin SR~\cite{Hagelstein:2015egb} \\
\hline
$\gamma_{E1 E1}$  
& \quad $-4.3$ \quad \quad \, ($10^{-4}$ fm$^4$) \quad
& DR~\cite{Holstein:1999uu} \\
\hline
$\gamma_{M1 M1}$  
& \quad $\hpm 2.9$ \quad \quad \, ($10^{-4}$ fm$^4$) \quad
& DR~\cite{Holstein:1999uu} \\
\hline
$\gamma_{E1 M2}$ 
& \quad $-0.1$ \quad \quad \, ($10^{-4}$ fm$^4$) \quad
& DR~\cite{Holstein:1999uu}  \\
\hline
$\gamma_{M1 E2}$  
& \quad $\hpm 2.1$ \quad \quad \, ($10^{-4}$ fm$^4$) \quad
& DR~\cite{Holstein:1999uu} \\
\hline
\quad $\delta_{LT}$   
&  \quad  $\hpm 1.34$  \quad \quad ($10^{-4}$ fm$^4$) \quad  
& MAID2007~\cite{Drechsel:2007if}   \\
\hline
\end{tabular}\par
}
\caption{Empirical values for the polarizabilities used in estimating the recoil terms in Eqs.~(\ref{eq:mp12}) and (\ref{eq:mp21}). 
}
\label{tab:recoil}
\end{table}

The uncertainty bands on the BChPT curves are calculated as detailed
in Ref.~\cite{Lensky:2016nui} and represent a conservative estimate
of corrections due to higher orders in the chiral expansion.
On the other hand, one can see that the use of the form factor
in the $\gamma N\Delta$ vertex is an important part of the presented
result. To estimate the uncertainty due to the form factor,
one notes that empirical data on electromagnetic nucleon-Delta transitions at low $Q^2$
allow one to extract the form factor with a precision
of the order of $2\%$; see, e.g., Ref.~\cite{Joo:2001tw}.
Varying the form factor within this range would result in changes of $M_1^{(2)}(Q^2)$ and $M_2^{(1)}(Q^2)$ at least an order of magnitude smaller than the shown uncertainty bands. We thus neglect the
uncertainty due to this source, expecting that any form factor
that describes electromagnetic nucleon-Delta transitions reasonably
well should give results close to those presented here.

The arguments concerning the uncertainty estimate also apply to the subtraction function $\bar{T}_1(0,Q^2)$;
see a more detailed discussion thereof in \secref{Subtraction}.

Finally, we can also extract an empirical estimate for the longitudinal polarizability in Eq.~(\ref{alphaL}).
For the term $M^{(4)}_{1}(0)$, we use the empirical sum-rule evaluation of Eq.~(\ref{eq:higherbaldin}) yielding~\cite{Gryniuk:2015eza}: 
$M^{(4)}_{1}(0) = 6.0 \cdot 10^{-4} \mathrm{fm}^{5}$. 
Using the BC fit values for $M_1^{(2)\prime}(0)$ and $M_2^{(1)\prime}(0)$, listed in Tables~\ref{tab:sr1} and \ref{tab:sr2}, we then obtain an empirical estimate for $\alpha_L$:
\bea
\alpha_L \simeq 1.1 \cdot 10^{-4}~\mathrm{fm}^5.\eqlab{alphaLprediction}
\eea
This polarizability has been calculated in BChPT at NLO~\cite{Lensky:2014dda}:~$\alpha_L \simeq 2.3 \cdot 10^{-4}$ fm$^5$.  We have checked that the same value is obtained by evaluating  
the separate BChPT contributions in Eq.~(\ref{alphaL}).

\section{Low-$\boldsymbol{Q}$ behavior of the Subtraction Function }\seclab{Subtraction}

In this section, we study the $Q^2$ dependence of the  subtraction function, $\bar{T}_1(0,Q^2)$, which is of interest for the (muonic) hydrogen Lamb shift calculations. It is the part of the TPE correction in the lepton-proton system noncalculable through the sum rules. In what follows, we will verify the analyticity constraint derived in Eq.~\eqref{eq:T1pp} and give estimates for the low-energy coefficient $b_{3,0}$. As a result, one constrains the subtraction contribution to the Lamb shift.

The LEX given in Eq.~\eqref{eq:T1pp} relates the second derivative of the subtraction function, $\bar{T}_1^{\prime\prime}(0)$, to scalar and spin polarizabilites known from RCS, the GP slope $\beta^\prime_{M1}$ known from VCS, and the low-energy coefficient $b_{3,0}$. Analogously to \secref{Verifications}, we verify \Eqref{T1pp} with the Delta-exchange graph contribution at ${\mathcal{O}}(p^4/\varDelta)$ in BChPT. As explained earlier, the validity of the constraint is not affected by adding a dipole form factor dependence to the magnetic coupling $g_M$ or, in general, by the inclusion of an arbitrary $Q^2$ dependence of the $\gamma N \Delta$ couplings. Once the constraint is verified, it can be used to make a prediction for $b_{3,0}$ at NLO in BChPT. As before, we rely on the results previously derived in Refs.~\cite{Lensky:2014dda,Lensky:2009uv,Lensky:2015awa,Lensky:2016nui}. The corresponding BChPT values [again, with
the use of the form factor in the Delta pole, as given by Eq.~\eqref{eq:DipoleGM}], as well as empirical and dispersive estimates
of all quantities entering Eq.~\eqref{eq:T1pp}, are given in Table~\ref{tab:subtraction}.

\begin{figure}[h]
\begin{center}
\includegraphics[width=0.7\textwidth]{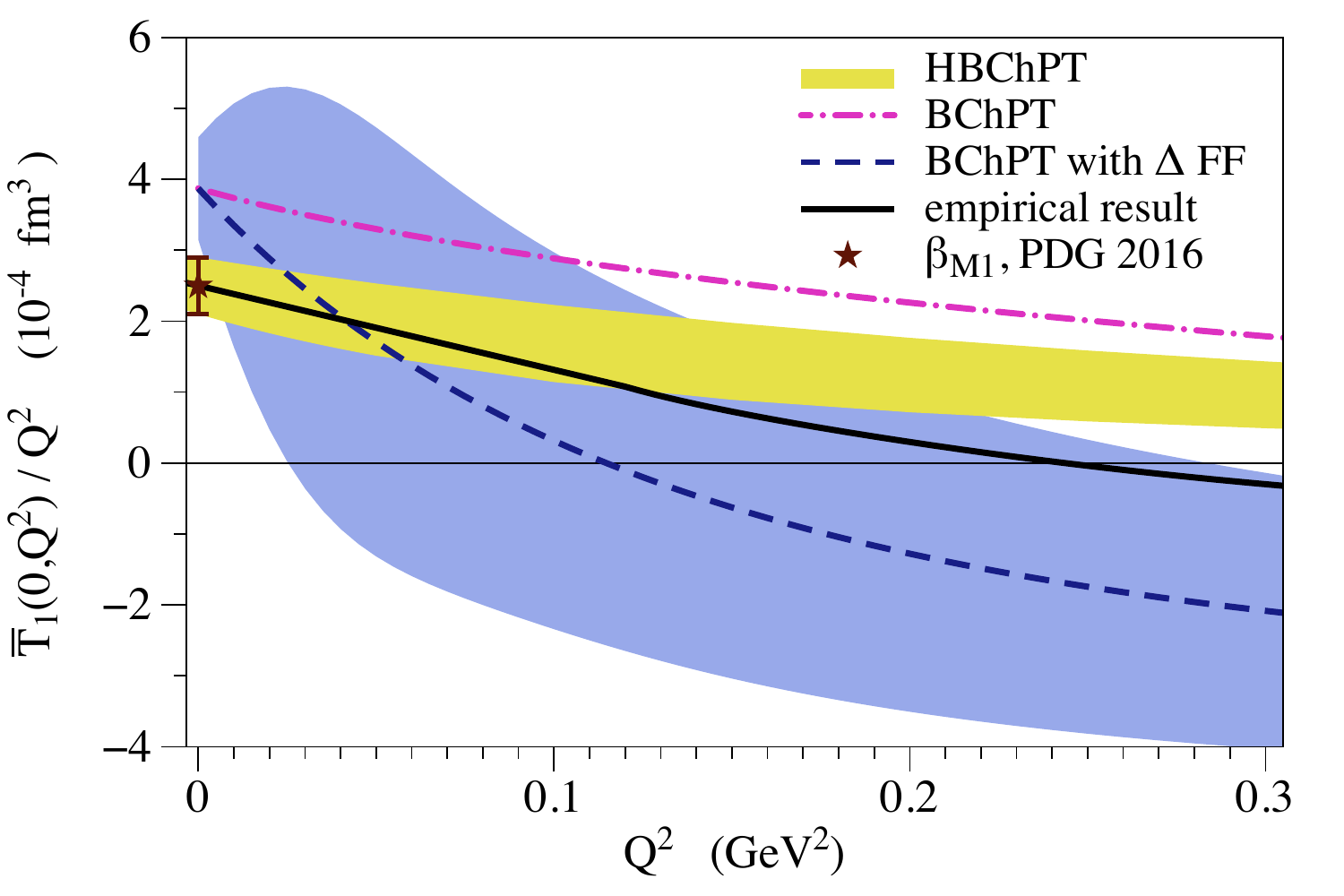}
\caption{The low-$Q^2$ behavior of the non-Born piece of the subtraction function. Shown are: the HBChPT calculation~\cite{Birse:2012eb} (dark yellow band), 
the BChPT calculation of this work (blue dashed and magenta dashed-dotted curves show the results with and without the form factor, respectively;
the wider blue band shows the uncertainty
of the BChPT result with the form factor, estimated in Ref.~\cite{Lensky:2016nui}), 
and the empirical superconvergence relation estimate of Ref.~\cite{Tomalak:2015hva} (black solid curve). 
At the real photon point, the PDG 2016 value of $\beta_{M1}=(2.5\pm 0.4)\times 10^{-4}~\text{fm}^3$~\cite{Olive:2016xmw} is shown. Note that the HBChPT curve is shifted to reproduce that value, whereas Ref.~\cite{Birse:2012eb} uses a larger value $\beta_{M1}=(3.15\pm 0.50)\times 10^{-4}~\text{fm}^3$ found in the most recent
HBChPT fit~\cite{McGovern:2012ew}.
}
\label{fig:subtraction}
\end{center}
\end{figure}

\begin{table}[h]
\begin{tabular}{c|c||c|c|c|c}
\hline\hline
Source & $\frac{1}{2}\bar{T}_1^{\prime\prime}(0)$ & $\alpha_\mathrm{em} b_{3,0}$ & $\beta_{M2}/6$ & $2 \beta_{M1}^\prime$
& $1/M^2$ recoil 
\\
\hline
$\pi N$ loops
& $-0.06$ & $\hpm 0.001$ & $-1.40$ & $\hpm 1.36$ & $-0.02$\\
%\hline
$\pi \Delta$ loops
& $-0.10$ & $    -0.005$ & $-0.44$ & $\hpm 0.37$ & $-0.02$\\
%\hline
$\Delta$ exchange
& $-1.98$ & $\hpm 0.11$ & $-0.75$ & $-1.42$ & $\hpm 0.08$ \\
\hline
Total
& $-2.14 \pm 0.98$ & $\hpm 0.11 \pm 0.05$ & $-2.59 \pm 0.59$ & $\hpm 0.31 \pm 0.50$ & $\hpm 0.04\pm 0.01$\\
\hline\hline
Empirical
& $-0.47$ & $3.96$   & $-4.10$ & $-0.36$ & $\hpm 0.03$\\
& \quad \quad estimate \cite{Tomalak:2015hva} \quad \quad
& Eq.~(\ref{b30Formula})
& \quad DR~\cite{Holstein:1999uu} \quad 
& DR~\cite{Pasquini:2001yy, Drechsel:2007if}
& PDG 2016~\cite{Olive:2016xmw}  \\
\hline\hline
\end{tabular}
\caption{Values of the low-energy coefficients entering the $Q^4$ term of the subtraction function $\bar T_1(0,Q^2)$, given by Eq.~(\ref{eq:T1pp}). All quantities are given in units of $10^{-4}~\text{fm}^5$. The first four rows are different contributions in BChPT: the $\Delta$-exchange contributions serve as a verification of the LEX constraint.
Errors are estimated as detailed in Ref.~\cite{Lensky:2015awa}.
The last row corresponds with empirical results either from dispersive (DR) estimates or the Particle Data Group (PDG). 
The value of $b_{3,0}$ in the last row is obtained from Eq.~(\ref{b30Formula})
by using the other values in that row as input.  
\label{tab:subtraction}
}
\end{table}

It is interesting to note that the value of $b_{3,0}$ obtained in BChPT turns out to be rather small
compared to other quantities entering Eq.~\eqref{eq:T1pp} and is driven by the Delta-exchange graph, with $\pi N$ and $\pi \Delta$ loops giving negligible contributions. The smallness of the
$\pi N$- and $\pi\Delta$-loop terms in $b_{3,0}$ could be considered accidental, given that it results from very efficient cancellations between
the different terms in Eq.~(\ref{eq:T1pp}).

Let us now compare the behavior of the subtraction function  
in different approaches. In Fig.~\ref{fig:subtraction}, we show $\bar{T}_1(0,Q^2)/Q^2$ as obtained in
BChPT and heavy-baryon chiral perturbation theory (HBChPT)~\cite{Birse:2012eb} (note that
the latter calculation uses a dipole form factor [with the slope
matched to the HBChPT expansion at low $Q^2$] to model
the large-$Q^2$ behavior of the subtraction function) and an estimate from the superconvergence relation~\cite{Tomalak:2015hva}. At the real photon point, $\bar{T}_1(0,Q^2)/Q^2$ is given by the magnetic dipole polarizability $\beta_{M1}$, cf.\ \Eqref{T1nb}. The figure shows that the BChPT curve with no $\gamma N\Delta$
form factor is close to the HBChPT one; note that the static value in the latter curve was fixed to the PDG value of $\beta_{M1}=(2.5\pm 0.4)\times 10^{-4}~\text{fm}^3$~\cite{Olive:2016xmw} rather than the larger value $\beta_{M1}=(3.15\pm 0.50)\times 10^{-4}~\text{fm}^3$
(which is typical of modern HBChPT~\cite{McGovern:2012ew} and BChPT~\cite{Lensky:2014efa} fits), used in Ref.~\cite{Birse:2012eb}.
The form factor on the magnetic $\gamma N\Delta$ coupling
increases the (negative) slope of the subtraction function at $Q^2=0$, 
as can be seen from Table~\ref{tab:subtraction} by
comparing the BChPT result (with form factor) with the empirical estimate.
It suppresses
the Delta-exchange contribution to the subtraction function at nonzero $Q^2$, and since the
$\pi N$- and $\pi\Delta$- loop contributions are negative, the result  with the form factor shows a zero crossing in the broad $Q^2$ range between $0.05$ and $0.25$~GeV$^2$.

  \begin{figure}[h!] 
    \centering 
       \includegraphics[width=4.5cm]{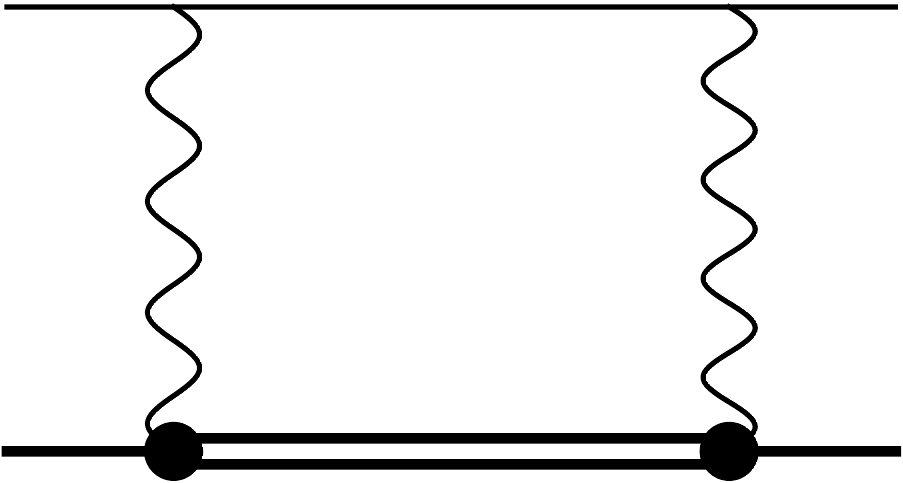}
       \caption{Two-photon-exchange diagram with intermediate $\Delta(1232)$ excitation.}
              \label{fig:TPEDelta}
\end{figure}

Let us now turn to the contribution of the subtraction term in the TPE correction to the Lamb shift in $\mu$H and in particular to the effect of the $\Delta(1232)$ excitation, shown in \Figref{TPEDelta}.
As seen in Fig.~\ref{fig:subtraction}, the subtraction function
changes a lot depending on the treatment of the Delta-exchange
contribution. However, as argued in Ref.~\cite{Alarcon:2013cba},
the total contribution of the Delta exchange to the Lamb shift
in $\mu$H turns out to be rather small due to cancellations
between the subtraction and inelastic terms. This picture as well as
the value of the total Delta-exchange contribution only
very weakly depend on the parametrization of the $\gamma N\Delta$
transition. We will demonstrate it in detail below; for this purpose, we briefly recall the TPE formalism (see, e.g., Ref.~\cite{Carlson:2011zd}).
The $n$th $S$-level shift in the (muonic) hydrogen spectrum due to {\it forward} TPE is related to the spin-independent {\it forward} VVCS amplitudes,
\be
\Delta E^\mathrm{TPE}(nS)= 8\pi e^2 m \,\phi_n^2\,
\frac{1}{i}\int_{-\infty}^\infty \!\frac{\dd\nu}{2\pi} \int \!\!\frac{\dd \bq}{(2\pi)^3}   \frac{\left(Q^2-2\nu^2\right)T_1(\nu,Q^2)-(Q^2+\nu^2)\,T_2(\nu,Q^2)}{Q^4(Q^4-4m^2\nu^2)},\eqlab{VVCS_LS}
\ee
where $m$ is the lepton mass, $\phi_n^2=1/(\pi n^3 a^3)$ is the wave function at the origin, $a^{-1}=\al_\mathrm{em} m_r$ is the inverse Bohr radius and $m_r$ is the reduced mass of the lepton-proton system. 
Recall also that the Lamb shift is the difference between the shifts
of the $2P$ and $2S$ levels; the TPE contribution to the former
is negligible, and the TPE contribution to the Lamb shift is thus
just $-\Delta E^\text{TPE}(2S)$.
Obviously, the polarizability effect on the hydrogen spectrum is described by the non-Born amplitudes $\bar{T}_1$ and $\bar{T}_2$. This effect can be split into the contribution of the subtraction function $\bar{T}_1(0,Q^2)$,
\be
\Delta E^{\mathrm{subtr.}}(nS)=\frac{2 e^2 m\,\phi_n^2}{\pi}\,\int_0^\infty \frac{\dd Q}{Q^3}\frac{v_l+2}{(1+v_l)^2}\, \bar{T}_1(0,Q^2),\eqlab{subpol}
\ee
with $v_l = \sqrt{1+4m^2/Q^2}$, and contributions of the inelastic structure functions (Ref.~\cite{Hagelstein:2015egb}, Sec.\ 6):
\bea
\Delta E^{\mathrm{inel.}}(nS)&=&-32\alpha_\mathrm{em}^2 Mm\,\phi_n^2\,\int_0^\infty \frac{\dd Q}{Q^5}\,\int_0^{x_0}\dd x\frac{1}{(1+v_l)(1+\sqrt{1+x^2\tau^{-1}})}\eqlab{inelasticpol}\\
&&\times\Bigg\{\left[1+\frac{v_l\sqrt{1+x^2\tau^{-1}}}{v_l+\sqrt{1+x^2\tau^{-1}}}\right]F_2(x,Q^2)\nn\\
&&+\frac{2x}{(1+v_l)(1+\sqrt{1+x^2\tau^{-1}})}\left[2+\frac{3+v_l\sqrt{1+x^2\tau^{-1}}}{v_l+\sqrt{1+x^2\tau^{-1}}}\right]F_1(x,Q^2)\Bigg\},\qquad\nn
\eea
where $\tau=Q^2/(4M^2)$. The $\Delta(1232)$-exchange contribution to the $\bar{T}_1(0,Q^2)$ subtraction function reads~\cite{Hagelstein2017}
\bea
\bar{T}_1(0,Q^2)&=&\frac{\alpha_\mathrm{em} Q^4}{M_\Delta M_+ \omega_+}\left[\frac{g_M^2}{Q^2}+\frac{g_M g_E}{M M_+}-\frac{g_E^2 \varDelta}{M^2 M_+}\right.\eqlab{T1su}\\
&&\left.+\frac{g_M g_C}{M M_+}+\frac{2 g_E g_C \left(M\varDelta +Q^2\right)}{M^2 M_\Delta M_+}-\frac{g_C^2 \varDelta \left(M^2-Q^2\right)}{M^2 M_\Delta^2 M_+}\right],\nn
\eea
with $M_+=M_\Delta+M$ and $\omega_+=(M_\Delta^2-M^2+ Q^2)/2M_\Delta
$. Here, the second row contains terms proportional to the subleading Coulomb coupling $g_C$.

In Table \ref{tab:deltaLS}, we show the effect of TPE with intermediate $\Delta(1232)$ excitation on the $2S$ level in $\mu$H.\footnote{Note that the structure functions not only contain the $\Delta$ production, i.e., terms proportional to $\delta(x-x_\Delta)$ with $x_\Delta=\frac{Q^2}{M_\Delta^2-M^2+Q^2}$, but also contain terms proportional to $\delta(x)$.} 
As mentioned above, the magnetic coupling can be multiplied by a dipole form factor in order to model a vector-meson type of dependence;
the use of the form factor is specified in the table. For the prediction in the last row, the $\gamma N\Delta$ couplings were replaced by the Jones-Scadron nucleon-to-Delta transition form factors (see Ref.~\cite{Hagelstein2017} for the details of the calculation). These transition form factors were related to nucleon form factors by the finite-momentum transfer extension of their large-$N_c$ limit~\cite{Pascalutsa:2007wz}. The nucleon form factors were in turn described by an empirical parametrization~\cite{Bradford:2006yz}. As one can see from the table, the relatively large contribution of the subtraction function, $\bar{T}_1(0,Q^2)$ (second column), is largely cancelled by the contributions of the inelastic structure functions, $F_1$ and $F_2$ (third and fifth columns). The total effect of the $\Delta$(1232) resonance on the shift of the $2S$ state in $\mu$H is small~\cite{Hagelstein2017} (quoting the calculation with the Jones-Scadron form factors),
\begin{equation}
\Delta E^{\langle\Delta\text{-excit.}\rangle\,\mathrm{pol.}}(2S, \mu \text{H})=0.95\pm0.95\, \upmu\mathrm{eV}\;\eqlab{DeExLS},
\end{equation}
compared to the leading effect of  chiral dynamics~\cite{Alarcon:2013cba},
\begin{equation}
\Delta E^{\langle\text{LO}\rangle\,\mathrm{pol.}}(2S, \mu \text{H})=-8.2_{-2.5}^{+1.2}\, \upmu\mathrm{eV}\;.
\end{equation}
At the same time, a calculation of the TPE with $\Delta(1232)$ excitation, employing again Jones-Scadron form factors, allows for a meaningful prediction of the contribution of the subtraction term (i.e., a prediction independent from its combination with the inelastic contribution into the polarizability contribution, cf.\ the discussion in Ref.~\cite{Alarcon:2013cba}, Sec.~3) to the shift of the $2S$ state at LO plus $\Delta$ in BChPT,
\begin{equation}
\Delta E^{\langle\text{LO+$\Delta$}\rangle\,\text{subtr.}}(2S, \mu \text{H})=4.6_{-2.4}^{+2.3}\, \upmu\mathrm{eV},
\end{equation}
which is in good agreement with dispersive predictions \cite{Carlson:2011zd,Birse:2012eb}. Table~\ref{tab:LS}
shows a comparison of separate contributions to $\Delta E^\text{TPE}(2S)$ in different frameworks.\footnote{A different HBChPT prediction of the subtraction term that does not use form factors to model the high-$Q^2$ dependence and
includes the leading and subleading $\pi N$ and  $\pi\Delta$ loops, respectively,  can be found in Ref.~\cite{Peset:2014jxa}.}

To conclude this section, we note that ChPT here is an example which satisfies the sum rules. However, the hope is that the sum rules will provide a data-driven evaluation, independent of ChPT. For that, one would need to have an experimental determination of the constant $b_{3,0}$, which can become possible in
future doubly virtual Compton scattering measurements.

\begin{table} [h]
\centering
\begin{small}
\centering
\begin{tabular}{c||c|c||c|c||c}
\hline
$\Delta E(2S)$ from: &$\bar{T}_1(0,Q^2)$&$F_1(x,Q^2)$&$\bar{T}_1(\nu,Q^2)$&$\bar{T}_2(\nu,Q^2)$&Total \\ &\Eqref{subpol}&\Eqref{inelasticpol}&\Eqref{VVCS_LS}&\Eqref{VVCS_LS}&\Eqref{VVCS_LS} \\
\hline
\hline
$g_M$ (without dipole FF) &$13.19$&$-4.31$&$8.88$&$-7.38$&$1.50$\\
\hline
$g_M$ (with dipole FF) &$8.01$&$-1.99$&$6.02$&$-5.10$&$0.92$\\
\hline
$G_M^*, G_E^*, G_C^*$ (Jones-Scadron FFs) &$7.58$&$-1.82$&$5.76$&$-4.82$&$0.95$\\
\hline
\end{tabular}
\caption{Contribution of the $\Delta(1232)$ excitation to the $2S$-level shift in $\mu$H. All values are given in $\upmu\mathrm{eV}$. For the dipole form factor (FF), we use $\Lambda^2=0.71~\text{GeV}^2$. In the last row, we use the empirical parametrization  \cite{Bradford:2006yz}.\label{tab:deltaLS}}
\end{small}
\end{table}

\begin{table}[h]
\label{tab:LS}
\begin{tabular}{l||r|c|c}
\hline
             & DR/HBChPT & BChPT (LO)~\cite{Alarcon:2013cba} & BChPT (LO + $\Delta$) \\
\hline\hline
$\Delta E^\text{inel}(2S)$  & $-12.7\pm 0.5$~\cite{Carlson:2011zd} & $-5.2$
& $-11.8^{+2.1}_{-2.5}$         \\
$\Delta E^\text{subtr}(2S)$ & $\hphantom{-1}4.2\pm 1.0$~\cite{Birse:2012eb}  &   $-3.0$       &   $\hphantom{-}4.6^{+2.3}_{-2.4}$       \\
\hline
$\Delta E^\text{pol}(2S)$   & $\hphantom{1}-8.5\pm 1.1$~\cite{Antognini:2013jkc}    &   $-8.2_{-2.5}^{+1.2}$   &  $-7.3^{+1.5}_{-2.7}$\\
\hline
\end{tabular}
\caption{$\Delta E^\text{TPE}(2S)$ contributions in different
calculations, all given in $\upmu$eV. The last line is the sum of inelastic (inel) and subtraction (subtr) contributions.}
\end{table}

\section{Conclusions}\seclab{Conclusions}
The main result of this work is given by the VVCS sum rules in Eqs.~\eref{mp12} and \eref{mp21}, and the LEX constraint in~\Eqref{T1pp}. For the derivation, the known CS formalism, reviewed in the beginning of~\secref{Formalism}, was used. 
At second order in energy ($\nu^2$) or momentum transfer ($Q^2$), the unpolarized nucleon response in the CS process is fully described in terms of electric and magnetic dipole polarizabilities. In this work, we have fully quantified the response of the double virtual CS to fourth order, including terms in $\nu^4, \nu^2 Q^2$, and $Q^4$. 
The new forward sum rules we have derived establish relations between RCS, VCS, and VVCS observables at this order. In particular, they give access to the VVCS low-energy coefficients $b_{4,1}$ and $b_{19,0}$ through moments of the nucleon structure functions, VCS GPs, and static scalar and spin polarizabilities; see Eqs.~\eqref{b4} and \eqref{b19}, respectively. 
From a practical point of view, this is an important result because $b_{19,0}$ does not appear in RCS or VCS experiments, and an empirical extraction of $b_{4,1}$ from VCS would be difficult due to higher-order GPs. The sum rule involving the low-energy coefficient $b_{4,1}$ was verified with the full NLO BChPT calculation, where all quantities entering Eq.~\eqref{eq:mp12} were calculated independently from the different CS processes.
The other sum rule and the LEX constraint were verified with the $\Delta$-exchange graph contribution at ${\mathcal{O}}(p^4/\varDelta)$ in BChPT. The theoretical and empirical results for the moments of proton structure functions $M_1^{(2)}$ and $M_2^{(1)}$, cf. Figs.~\ref{fig:momentsf1} and~\ref{fig:momentsf2},  and for most low-energy constants entering the two newly established sum rules, cf. Tables~\ref{tab:sr1} and~\ref{tab:sr2}, were found to be in reasonable good agreement. 

The remaining unknown in the doubly virtual CS process at order $Q^4$ results from the low-energy coefficient $b_{3,0}$, which enters the VVCS subtraction function $\bar{T}_1(0,Q^2)$. 
The latter is also the main hadronic uncertainty in the estimate of the TPE correction to the muonic-hydrogen Lamb shift. 
Our NLO BChPT calculation yields a very small value for $b_{3,0}$. We have shown that this result originates predominantly from the $\Delta$-pole contribution. 
The corresponding NLO BChPT prediction of the subtraction function displays a sign change induced by the form factor dependence of the $\Delta$-exchange graph. The LO plus $\Delta$ BChPT prediction for the polarizability contribution (subtraction term and inelastic term) to the $\mu$H Lamb shift is found to be in good agreement with dispersive calculations. Studying in particular the TPE with intermediate $\Delta$ excitation, we have shown that the sizeable contribution of the subtraction term is largely cancelled by the inelastic contribution, leading to a small polarizability effect of the $\Delta(1232)$ in the $\mu$H Lamb shift. 

To check the smallness of the low-energy coefficient $b_{3,0}$, 
as predicted by our NLO BChPT calculation, we noted that there is at present no direct experimental access to the slope of the VVCS subtraction function. 
In order to have some empirical guidance, we compared our BChPT result with the estimate based on a superconvergence relation~\cite{Tomalak:2015hva}. 
The latter yields a much smaller value (in absolute size) for the $Q^4$ term in the subtraction function $\bar{T}_1(0,Q^2)$, which then yields a significantly larger value for $b_{3,0}$. The superconvergence estimate of 
Ref.~\cite{Tomalak:2015hva} at lower values of $Q^2 \lesssim 1$~GeV$^2$ is constrained by existing nucleon structure function data in the resonance region ($W < 3$~GeV) as well as by HERA data at high energies ($W > 10$~GeV). 
However, in the intermediate $W$ region  
($3 \lesssim W \lesssim 10$~GeV) at finite $Q^2$, the empirical estimate is quite uncertain because of the scarce data situation in that region. Forthcoming structure function data from the JLab 12 GeV facility will allow us to further improve such superconvergence relation estimates for $b_{3,0}$. It may also be very worthwhile to directly access $b_{3,0}$ through a low-energy doubly virtual CS experiment. The formalism laid out in the present work provides the unpolarized hadronic tensor entering the description of such a process. We leave the study of the doubly virtual CS observables necessary to measure the low-energy coefficient $b_{3,0}$ as a topic for future work.

\acknowledgments

This work was supported by the Deutsche Forschungsgemeinschaft (DFG) 
in part through the Collaborative Research Center [The Low-Energy Frontier of the Standard Model (SFB 1044)], and in part through the Cluster of Excellence [Precision Physics, Fundamental
Interactions and Structure of Matter (PRISMA)]. The work of F.\ H.\ is supported by the Swiss National Science Foundation.

\appendix
\section{RCS limit}
\label{RCS}

In this Appendix, we discuss, as special case of the doubly virtual Compton process, the RCS limit, corresponding with $q^2 = q^{\prime \, 2} = 0$. 
The spin-independent part of the RCS amplitude is described by
\bea
\label{rcs}
M^{\mu \nu}(\mathrm{RCS}) \big|_{\mathrm{spin \, indep.}}&=&
 \left( g^{\mu\nu}-\frac{q^{\prime \mu}q^{\nu}}{q \cdot q^\prime}\right) 
 \left\{ - q\cdot q^\prime \, B_1(0,0,q \cdot q^\prime, M \nu) 
 - (2 M \nu)^2 \, B_2(0,0,q \cdot q^\prime, M \nu)  \right\}
\nonumber \\
&- & 4 q \cdot q^\prime \left( P^{\mu} - \frac{M \nu}{q \cdot q^\prime} q^{\prime \mu} \right)   
\left( P^{\nu} - \frac{M \nu}{q \cdot q^\prime} q^{\nu} \right)  B_2(0,0,q \cdot q^\prime, M \nu) ,
\eea
as the other three tensors in Eq.~(\ref{eq:dvcstensors}) do not contribute to the RCS limit.

Both a dispersive formulation as well as a LEX for RCS is conventionally described by an equivalent set of  amplitudes $A_i(\nu, t)$, for $i = 1,...,6$, free of kinematic singularities and constraints, see Refs.~\cite{Lvov:1980wp,Lvov:1996rmi}, which can be obtained as linear combinations of the $B_i$ amplitudes. 
We give here explicitly the relations between the amplitudes $B_1$ and $B_2$, which appear in Eq.~(\ref{rcs}), and the $A_i$ amplitudes~\cite{Pasquini:2001yy},
\bea
B_1(0,0,q \cdot q^\prime, M \nu) &=& \frac{1}{4 \pi \alpha_\mathrm{em}} 
\left\{ A_1(\nu, t) - A_3(\nu, t) - A_6(\nu, t) + \frac{t}{4 M^2} A_3(\nu, t) - \frac{\nu^2}{M^2} A_4(\nu, t) \right\}, 
\label{b1rcs} \\
B_2(0,0,q \cdot q^\prime, M \nu) &=& \frac{1}{4 \pi \alpha_\mathrm{em}} \frac{1}{2 M^2}
\left\{ A_3(\nu, t) + A_6(\nu, t) - \frac{t}{4 M^2} A_4(\nu, t) \right\} ,
\label{b2rcs} 
\eea
with $t = - 2 q \cdot q^\prime$ for the RCS process. 

The LEX of the non-Born parts of the amplitudes $A_i$ can be written as~\cite{Babusci:1998ww,Holstein:1999uu,Drechsel:2002ar}
\bea
\bar{A}_i(\nu, t) = a_i + a_{i, \nu} \, \nu^2 + a_{i, t} \, t +  {\cal O}(k^4), \quad \quad i = 1,...,6,
\eea 
where $k^4$ stands for higher-order terms either in $\nu^4$, $\nu^2 t$, or $t^2$.  The low-energy coefficients at zeroth order, $a_i$, can be expressed in terms of nucleon scalar dipole and lowest-order spin polarizabilities, 
whereas the low-energy coefficients at second order,  $a_{i, \nu}$ and $a_{i, t}$, have been worked out 
in terms of quadrupole, dispersive, or higher-order spin polarizabilities~\cite{Babusci:1998ww,Holstein:1999uu,Drechsel:2002ar}. For an example, we quote here the expressions for the combinations of the lowest-order coefficients which enter the LEXs for the amplitudes $B_1$ and $B_2$ of Eqs.~(\ref{b1rcs}) and (\ref{b2rcs}),
\bea
a_1 - a_3 - a_6 &=&  4 \pi  \beta_{M1}, \nonumber \\
a_3 + a_6 &=& - 2  \pi  \left( \alpha_{E1} + \beta_{M1} \right), 
\eea
in terms of the nucleon electric ($\alpha_{E1}$) and magnetic ($\beta_{M1}$)  dipole polarizabilities. 
The detailed expressions for all coefficients $a_i$, $a_{i, t}$, and $a_{i, \nu}$ can be found in 
Ref.~\cite{Holstein:1999uu}. 
In terms of these low-energy coefficients, we can then construct  the LEXs of the non-Born parts of the amplitudes $B_1$ and $B_2$ of Eqs.~(\ref{b1rcs}) and (\ref{b2rcs}) in the RCS limit as
\bea
\bar{B}_1(0,0,q \cdot q^\prime, M \nu) &=& \frac{1}{4 \pi \alpha_\mathrm{em}} 
\left\{ a_1 - a_3 - a_6 
- 2 \left[ a_{1, t} - a_{3, t} - a_{6, t} + \frac{a_3}{4 M^2} \right] q \cdot q^\prime \right. \nonumber \\
&&\left. \hspace{3.9cm} +  \left[ a_{1, \nu} - a_{3, \nu} - a_{6, \nu} - \frac{a_4}{M^2} \right]  \nu^2   
\right\} +  {\cal O}(k^4),  
\label{b1rcslex} \\
\bar{B}_2(0,0,q \cdot q^\prime, M \nu) &=& \frac{1}{4 \pi \alpha_\mathrm{em}} \frac{1}{2 M^2}
\left\{ a_3 + a_6 
- 2 \left[ a_{3, t} + a_{6, t} - \frac{a_4}{4 M^2} \right] q \cdot q^\prime  
+  \left[ a_{3, \nu} + a_{6, \nu}  \right]  \nu^2   
\right\} +  {\cal O}(k^4). \quad \; 
\label{b2rcslex} 
\eea
By substituting the relations between the low-energy coefficients 
$a_i$, $a_{i, t}$, and $a_{i, \nu}$ and the polarizabilities, 
the RCS process then allows us to determine the coefficients in the low-energy expansion given by Eq.~(\ref{lexdvcsa}) for the non-Born amplitudes $\bar B_1$ and $\bar B_2$. 
The corresponding expressions for these coefficients are given in Eqs.~(\ref{b10})-(\ref{b22c}).
Note that the recoil terms (proportional to $1/M$ and $1/M^2$) in 
Eqs.~(\ref{b10})-(\ref{b22c}) 
arise due to the transformation from the Breit frame, in which the polarizabilities such as $\beta_{M1}$, $\beta_{M1, \nu}$, $\beta_{M2}$, ...,
are defined, and the LEX of the Compton amplitude in terms of the $A_1,...,A_6$.

\section{VCS limit}
\label{VCS}

Another special limit of the doubly virtual Compton process 
is the nonforward VCS process, which corresponds with an outgoing real photon, i.e. $q^{\prime \, 2} = 0$, and an initial spacelike virtual photon with virtuality $q^2=-Q^2<0$. 
The VCS process  can generally be parametrized in terms of $12$ independent amplitudes, $f_i(q^2, q \cdot q', q \cdot P)$ for $i = 1,...,12$, 
as introduced in Ref.~\cite{Drechsel:1997xv}.  
The nucleon spin-independent VCS process is described by three amplitudes, which are related to the doubly virtual Compton amplitudes 
$B_i$ entering Eq.~(\ref{eq:dvcsunpol}) as
\begin{subequations}
\bea
B_1(q^2, 0, q \cdot q^\prime, q \cdot P) = f_1(q^2, q \cdot q^\prime, q \cdot P), \label{b1vcs} \\
B_2(q^2, 0, q \cdot q^\prime, q \cdot P) = f_2(q^2, q \cdot q^\prime, q \cdot P), \label{b2vcs} \\
B_4(q^2, 0, q \cdot q^\prime, q \cdot P) = f_3(q^2, q \cdot q^\prime, q \cdot P) \label{b4vcs} . 
\eea
\end{subequations}
Note that the remaining two amplitudes $B_3$ and $B_{19}$ which are needed 
to fully specify the spin-independent doubly virtual Compton amplitude of Eq.~(\ref{eq:dvcsunpol}) 
cannot be accessed in the VCS process, as the corresponding tensors decouple when the 
outgoing photon is real ($q^{\prime \, 2} = 0$). 

The VCS experiments at low outgoing photon energies 
can also be analyzed in terms of LEXs, as proposed in Ref.~\cite{Guichon:1995pu}. 
For this purpose, the VCS tensor has been split in Ref.~\cite{Guichon:1995pu} into a Born part, which is defined as the nucleon intermediate state contribution using the $\gamma^\ast N N$ vertex of Eq.~(\ref{emvertex}), and a non-Born part. The latter describes the response of the nucleon to the quasistatic electromagnetic field, due to the nucleon's internal structure. To obtain the lowest-order nucleon structure terms, one considers the response linear in the energy of the produced real photon.  
This linear response of the non-Born VCS tensor, 
i.e., the limit $q^\prime \to 0$ at arbitrary virtuality $Q^2$ of the initial photon, can be parametrized by six independent GPs \cite{Guichon:1995pu,Drechsel:1998zm}. 
The GPs can be accessed in experiment through the 
$e N \to e N \gamma$ process; see the reviews \cite{Guichon:1998xv,Drechsel:2002ar} for more details. 
At lowest order in the outgoing photon energy, there are two spin-independent GPs, denoted by 
$P^{(L1, L1)0}$ and $P^{(M1, M1)0}$, and four spin GPs, 
denoted by $P^{(L1, M2)1}$, $P^{(M1, L2)1}$, $P^{(L1, L1)1}$, and $P^{(M1, M1)1}$, 
which are all functions of $Q^2$.\footnote{Equivalently, they can be considered
as functions of the three-momentum $\bar {\rm q}$ of the virtual photon, which is conveniently defined in the c.m. system of the $\gamma^\ast  N$ system, and given by $\bar{\rm q}^2=Q^2(1+ \tau)$; this definition
is used in Ref.~\cite{Guichon:1995pu}.}
In this notation, $L$ stands for the longitudinal (or electric) and $M$ stands for the magnetic nature of the transition respectively.   
One usually defines the electric and magnetic GPs as
\begin{subequations}
\bea
\beta_{M1}(Q^2) &=& - \alpha_\mathrm{em} \sqrt{\frac{3}{8}} \, P^{(M1, M1)0}(Q^2), 
\label{gpbeta} \\
\alpha_{E1}(Q^2) &=& - \alpha_\mathrm{em} \sqrt{\frac{3}{2}} \, P^{(L1, L1)0}(Q^2),
\label{gpalpha} 
\eea
\end{subequations}
which are related to the RCS static polarizabilities as
\begin{equation}
\alpha_{E1}(0)=\alpha_{E1},\qquad \beta_{M1}(0)=\beta_{M1}.
\end{equation}
The GPs can be expressed  
in terms of the non-Born parts $\bar f_i$ of the invariant amplitudes $f_i$. 
Using the shorthand notation
\bea
\bar f_i(Q^2) \equiv \bar{f}_i(q^2 = -Q^2, 0, 0), 
\eea
the spin-independent magnetic and electric GPs can be, respectively, obtained as~\cite{Drechsel:1998zm}
\begin{subequations}
\bea
\beta_{M1}(Q^2)&=&\hphantom{-}
\alpha_\mathrm{em}\sqrt{\frac{1 + \tau}{1 + 2 \tau}} \, \bar{f}_1(Q^2),
\label{eq:PM1M10} \\
\alpha_{E1}(Q^2)&=&-\alpha_\mathrm{em} \sqrt{\frac{1 + \tau}{1 + 2 \tau}}
\left[\bar{f}_1(Q^2)+4M^2(1+\tau)\bar{f}_2(Q^2) 
+ Q^2 \left(2\bar{f}_6(Q^2)+\bar{f}_9(Q^2)-\bar{f}_{12}(Q^2)\right)
\right]. \qquad\; \label{eq:PL1L10} 
\eea
\end{subequations}
At $Q^2 = 0$, these relations reduce to 
\begin{subequations}
\bea
\bar{f}_1(0) &=& \hphantom{-}\frac{1}{\alpha_\mathrm{em} } \beta_{M1},  
\label{f10} \\
\bar{f}_2(0) &=&  - \frac{1}{\alpha_\mathrm{em} } \frac{1}{(2M)^2} \left( \alpha_{E1} + \beta_{M1} \right).
\label{f20}
\eea
\end{subequations} 
Using the relations of Eqs.~(\ref{f10}) and (\ref{f20}) 
as a limit of Eqs.~(\ref{b1vcs}) and (\ref{b2vcs}), one readily verifies the 
expressions obtained before in Eqs.~(\ref{b10}) and (\ref{b20}) for $b_{1, 0}$ and $b_{2,0}$ respectively. 
We can next consider the slopes at $Q^2 =0$ of the magnetic and electric GPs:
\begin{subequations}
\bea
\beta_{M1}^\prime &\equiv& \frac{\rm d\hphantom{Q^2}}{{\rm d} Q^2} 
\beta_{M1}(Q^2 )  \bigg|_{Q^2 =0}, 
 \label{GPslopeM} \\
\alpha_{E1}^\prime &\equiv& \frac{\rm d\hphantom{Q^2}}{{\rm{d}} Q^2} 
 \alpha_{E1}(Q^2 )  \bigg|_{Q^2 =0}.
 \label{GPslopeL}
 \eea
\end{subequations}
By taking the derivatives at $Q^2 = 0$ of Eqs.~(\ref{eq:PM1M10}) and (\ref{eq:PL1L10}), we obtain
\begin{subequations}
\bea
\beta_{M1}^\prime &=&\hphantom{-}
\alpha_\mathrm{em}  \, \left[ \bar{f}^\prime_1(0) - \frac{1}{8 M^2} \bar{f}_1(0)  \right],
\label{slopegpmagn} \\ 
\alpha_{E1}^\prime &=&-\alpha_\mathrm{em}
\left[\bar{f}^\prime_1(0) + 4M^2 \bar{f}^\prime_2(0) - \frac{1}{8 M^2} \bar{f}_1(0) + \frac{1}{2} \bar{f}_2(0) 
+ 2\bar{f}_6(0)+\bar{f}_9(0)-\bar{f}_{12}(0)
\right].
\label{slopegpel} 
\eea
\end{subequations}
The combination $2\bar{f}_6(0)+\bar{f}_9(0)-\bar{f}_{12}(0)$ in Eq.~(\ref{slopegpel}) 
can be expressed in terms of spin GPs 
using the expressions of Ref.~\cite{Drechsel:1998zm}. 
It was shown recently that a forward sum 
rule allows one to express this combination as~\cite{Pascalutsa:2014zna,Lensky:2017dlc}
\bea
2\bar{f}_6(0)+\bar{f}_9(0)-\bar{f}_{12}(0) = 
\frac{1}{\alpha_\mathrm{em}} \frac{1}{2 M} \, 
\left( - \delta_{LT} - \gamma_{E1E1} + \gamma_{E1M2} \right),
\label{f6912}
\eea
in terms of the RCS spin polarizabilities $\gamma_{E1E1}$ and $\gamma_{E1M2}$, as well as the longitudinal-transverse spin polarizability $\delta_{LT}$ at $Q^2 = 0$, 
which is accessed from a moment of the nucleon spin-dependent structure functions $g_1$ and $g_2$.

We can then determine two further low-energy coefficients as
\bea
b_{1, 2b} &=& -\bar f^\prime_1(0), \\ 
b_{2, 2b} &=& -\bar f^\prime_2(0). 
\eea
When using Eqs.~(\ref{slopegpmagn}) and (\ref{slopegpel}), 
we then obtain for the coefficients $b_{1, 2b}$ and $b_{2, 2b}$ 
the expressions of Eqs.~(\ref{b12b}) and (\ref{b22b}).


\begin{thebibliography}{99}


\bibitem{Guichon:1998xv} 
  P.~A.~M.~Guichon and M.~Vanderhaeghen,
  %``Virtual Compton scattering off the nucleon,''
  Prog.\ Part.\ Nucl.\ Phys.\  {\bf 41}, 125 (1998).
%  [hep-ph/9806305].
  %%CITATION = HEP-PH/9806305;%%


\bibitem{Drechsel:2002ar} 
  D.~Drechsel, B.~Pasquini, and M.~Vanderhaeghen,
  %``Dispersion relations in real and virtual Compton scattering,''
  Phys.\ Rep.\  {\bf 378}, 99 (2003).
%  [hep-ph/0212124].
  %%CITATION = HEP-PH/0212124;%%

\bibitem{Schumacher:2005an} 
  M.~Schumacher,
  %``Polarizability of the nucleon and Compton scattering,''
  Prog.\ Part.\ Nucl.\ Phys.\  {\bf 55}, 567 (2005). 
  %doi:10.1016/j.ppnp.2005.01.033
  %[hep-ph/0501167].
  %%CITATION = doi:10.1016/j.ppnp.2005.01.033;%%

 
%\cite{Holstein:2013kia}
\bibitem{Holstein:2013kia}
  B.~R.~Holstein and S.~Scherer,
 % ``Hadron Polarizabilities,''
  Ann.\ Rev.\ Nucl.\ Part.\ Sci.\  {\bf 64}, 51 (2014).
  %%doi:10.1146/annurev-nucl-102313-025555
%  [hep-ph/1401.0140].


\bibitem{Hagelstein:2015egb} 
  F.~Hagelstein, R.~Miskimen, and V.~Pascalutsa,
  %``Nucleon Polarizabilities: from Compton Scattering to Hydrogen Atom,''
  Prog.\ Part.\ Nucl.\ Phys.\  {\bf 88}, 29 (2016). 
 % doi:10.1016/j.ppnp.2015.12.001
 % [arXiv:1512.03765 [nucl-th]].
  %%CITATION = doi:10.1016/j.ppnp.2015.12.001;%%


\bibitem{GellMann:1954db} 
  M.~Gell-Mann, M.~L.~Goldberger, and W.~E.~Thirring,
  %``Use of causality conditions in quantum theory,''
  Phys.\ Rev.\  {\bf 95}, 1612 (1954).
  %doi:10.1103/PhysRev.95.1612
  %%CITATION = doi:10.1103/PhysRev.95.1612;%%
  
   \bibitem{Baldin}
A.~M.~Baldin, Nucl.\ Phys.\ {\bf 18}, 310 (1960).



\bibitem{Gerasimov:1965et} 
  S.~B.~Gerasimov,
  %``A Sum rule for magnetic moments and the damping of the nucleon magnetic moment in nuclei,''
  Yad.\ Fiz.\  {\bf 2}, 598 (1965)
  [Sov.\ J.\ Nucl.\ Phys.\  {\bf 2}, 430 (1966)].
  %%CITATION = SJNCA,2,430;%%

 
  

\bibitem{Drell:1966jv} 
  S.~D.~Drell and A.~C.~Hearn,
  %``Exact Sum Rule for Nucleon Magnetic Moments,''
  Phys.\ Rev.\ Lett.\  {\bf 16}, 908 (1966).
%  doi:10.1103/PhysRevLett.16.908
  %%CITATION = doi:10.1103/PhysRevLett.16.908;%%
 

 

\bibitem{Burkhardt:1970ti} 
  H.~Burkhardt and W.~N.~Cottingham,
  %``Sum rules for forward virtual Compton scattering,''
  Ann.\ Phys.\  {\bf 56}, 453 (1970).
  %doi:10.1016/0003-4916(70)90025-4
  %%CITATION = doi:10.1016/0003-4916(70)90025-4;%%
  
  %\cite{Schwinger:1975ti}
\bibitem{Schwinger:1975ti} 
  J.~S.~Schwinger,
  %``Source Theory Viewpoints in Deep Inelastic Scattering,''
  Proc.\ Nat.\ Acad.\ Sci.\  {\bf 72}, 1 (1975); {\bf 72}, 1559 (1975); Acta Phys.\ Austriaca Suppl.\  {\bf 14}, 471 (1975).
%  doi:10.1007/978-3-7091-8424-0_9, 10.1073/pnas.72.1.1
  %%CITATION = doi:10.1007/978-3-7091-8424-0_9, 10.1073/pnas.72.1.1;%%

%\cite{Cottingham:1963zz}
\bibitem{Cottingham:1963zz} 
  W.~N.~Cottingham,
  %``The neutron proton mass difference and electron scattering experiments,''
  Ann.\ Phys.\  {\bf 25}, 424 (1963).
  %doi:10.1016/0003-4916(63)90023-X
  %%CITATION = doi:10.1016/0003-4916(63)90023-X;%%
  %197 citations counted in INSPIRE as of 05 Feb 2018

%\cite{Gasser:1974wd}
\bibitem{Gasser:1974wd} 
  J.~Gasser and H.~Leutwyler,
  %``Implications of Scaling for the Proton - Neutron Mass - Difference,''
  Nucl.\ Phys.\ B {\bf 94}, 269 (1975).
%  doi:10.1016/0550-3213(75)90493-9
  %%CITATION = doi:10.1016/0550-3213(75)90493-9;%%
  %120 citations counted in INSPIRE as of 05 Feb 2018

%\cite{Gasser:2015dwa}
\bibitem{Gasser:2015dwa} 
  J.~Gasser, M.~Hoferichter, H.~Leutwyler, and A.~Rusetsky,
  %``Cottingham formula and nucleon polarisabilities,''
  Eur.\ Phys.\ J.\ C {\bf 75}, 375 (2015).
  %doi:10.1140/epjc/s10052-015-3580-9
  %[arXiv:1506.06747 [hep-ph]].
  %%CITATION = doi:10.1140/epjc/s10052-015-3580-9;%%
  %20 citations counted in INSPIRE as of 05 Feb 2018

\bibitem{Pascalutsa:2014zna} 
  V.~Pascalutsa and M.~Vanderhaeghen,
  %``Polarizability relations across real and virtual Compton scattering processes,''
  Phys.\ Rev.\ D {\bf 91}, 051503 (2015). 
%  doi:10.1103/PhysRevD.91.051503
%  [arXiv:1409.5236 [nucl-th]].
   
\bibitem{Lensky:2017dlc} 
  V.~Lensky, V.~Pascalutsa, M.~Vanderhaeghen, and C.~W.~Kao,
  %``Spin-dependent sum rules connecting real and virtual Compton scattering verified,''
   Phys.\ Rev.\ D {\bf 95}, 074001 (2017).
%   arXiv:1701.01947 [hep-ph].
  %%CITATION = ARXIV:1701.01947;%%

\bibitem{Kuhn:2008sy} 
  S.~E.~Kuhn, J.-P.~Chen, and E.~Leader,
  %``Spin Structure of the Nucleon - Status and Recent Results,''
  Prog.\ Part.\ Nucl.\ Phys.\  {\bf 63}, 1 (2009). 
  %[arXiv:0812.3535 [hep-ph]].
  %%CITATION = ARXIV:0812.3535;%%
 
\bibitem{Chen:2010qc} 
  J.~P.~Chen,
  %``Moments of Spin Structure Functions: Sum Rules and Polarizabilities,''
  Int.\ J.\ Mod.\ Phys.\ E {\bf 19}, 1893 (2010). 
%  [arXiv:1001.3898 [nucl-ex]].
  %%CITATION = ARXIV:1001.3898;%%


\bibitem{Martel:2014pba} 
  P.~P.~Martel {\it et al.} (A2 Collaboration),
  %``Measurements of Double-Polarized Compton Scattering Asymmetries and Extraction of the Proton Spin Polarizabilities,''
  Phys.\ Rev.\ Lett.\  {\bf 114}, 112501 (2015).
%  doi:10.1103/PhysRevLett.114.112501
%  [arXiv:1408.1576 [nucl-ex]].
  %%CITATION = doi:10.1103/PhysRevLett.114.112501;%%

\bibitem{d'Hose:2006xz}
  N.~d'Hose,
  %``Virtual Compton scattering at MAMI,''
  Eur.\ Phys.\ J.\ A {\bf 28S1}, 117 (2006).
%  doi:10.1140/epja/i2006-09-013-6
  %%CITATION = doi:10.1140/epja/i2006-09-013-6;%%
  %2 citations counted in INSPIRE as of 25 Oct 2016

\bibitem{Bourgeois:2006js}
  P.~Bourgeois {\it et al.},
  %``Measurements of the generalized electric and magnetic polarizabilities of the proton at low Q**2 using the VCS reaction,''
  Phys.\ Rev.\ Lett.\  {\bf 97}, 212001 (2006).
%  doi:10.1103/PhysRevLett.97.212001
%  [nucl-ex/0605009].
  %%CITATION = doi:10.1103/PhysRevLett.97.212001;%%


\bibitem{Bourgeois:2011zz}
  P.~Bourgeois {\it et al.},
  %``Measurements of the generalized electric and magnetic polarizabilities of the proton at low Q-2 using the virtual Compton scattering reaction,''
  Phys.\ Rev.\ C {\bf 84}, 035206 (2011).
%  doi:10.1103/PhysRevC.84.035206
  %%CITATION = doi:10.1103/PhysRevC.84.035206;%%

\bibitem{Correa:thesis}
L.~Correa, 
%{\it Measurement of the generalized polarizabilities of
%the proton by virtual Compton scattering at MAMI and $Q^2=0.2$~GeV${^2}$},
Ph.D.\ thesis, Johannes Gutenberg-Universit{\"a}t, Mainz and Universit{\'e} Blaise Pascal, Clermont-Ferrand, 2016.


\bibitem{Drechsel:1997xv} 
  D.~Drechsel, G.~Kn{\"o}chlein, A.~Y.~Korchin, A.~Metz, and S.~Scherer,
  %``Structure analysis of the virtual Compton scattering amplitude at low-energies,''
  Phys.\ Rev.\ C {\bf 57}, 941 (1998). 
  %doi:10.1103/PhysRevC.57.941
  %[nucl-th/9704064].
  %%CITATION = doi:10.1103/PhysRevC.57.941;%%



\bibitem{Tarrach:1975tu} 
  R.~Tarrach,
  %``Invariant Amplitudes for Virtual Compton Scattering Off Polarized Nucleons Free from Kinematical Singularities, Zeros and Constraints,''
  Nuovo Cimento\ A {\bf 28}, 409 (1975).
  %%CITATION = NUCIA,A28,409;%%


    


 \bibitem{Lensky:2016nui} 
  V.~Lensky, V.~Pascalutsa, and M.~Vanderhaeghen,
  %``Generalized polarizabilities of the nucleon in baryon chiral perturbation theory,''
  Eur.\ Phys.\ J.\ C {\bf 77}, 119 (2017).
 % doi:10.1140/epjc/s10052-017-4652-9
  %[arXiv:1612.08626 [hep-ph]].
  %%CITATION = doi:10.1140/epjc/s10052-017-4652-9;%%  


\bibitem{Hand:1963bb} 
  L.~N.~Hand,
  %``Experimental investigation of pion electroproduction,''
  Phys.\ Rev.\  {\bf 129}, 1834 (1963).
 % doi:10.1103/PhysRev.129.1834
  %%CITATION = doi:10.1103/PhysRev.129.1834;%%


 
\bibitem{Liang:2004tk} 
  Y.~Liang, M.~E.~Christy, R.~Ent, and C.~E.~Keppel,
  %``Q**2 evolution of generalized Baldin sum rule for the proton,''
  Phys.\ Rev.\ C {\bf 73}, 065201 (2006).
 % doi:10.1103/PhysRevC.73.065201
 % [nucl-ex/0410028].
  %%CITATION = doi:10.1103/PhysRevC.73.065201;%%  
  

    
    
\bibitem{Lensky:2014dda}
  V.~Lensky, J.~M.~Alarc\'on, and V.~Pascalutsa,
  %``Moments of nucleon structure functions at next-to-leading order in baryon chiral perturbation theory,''
  Phys.\ Rev.\ C {\bf 90}, 055202 (2014). 
%  [arXiv:1407.2574 [hep-ph]].
  %%CITATION = ARXIV:1407.2574;%%

\bibitem{Lensky:2009uv}
  V.~Lensky and V.~Pascalutsa,
  %``Predictive powers of chiral perturbation theory in Compton scattering off
  %protons,''
  Eur.\ Phys.\ J.\  C {\bf 65}, 195 (2010).
  %[arXiv:0907.0451 [hep-ph]].
  %%CITATION = EPHJA,C65,195;%%  

 
\bibitem{Lensky:2015awa} 
  V.~Lensky, J.~McGovern, and V.~Pascalutsa,
  %``Predictions of covariant chiral perturbation theory for nucleon polarisabilities and polarised Compton scattering,''
  Eur.\ Phys.\ J.\ C {\bf 75}, 604 (2015).
%  doi:10.1140/epjc/s10052-015-3791-0
%  [arXiv:1510.02794 [hep-ph]].
  %%CITATION = doi:10.1140/epjc/s10052-015-3791-0;%%
  
     \bibitem{Pascalutsa:2002pi} 
V.~Pascalutsa and D.~R.~Phillips,
    %``Effective theory of the Delta(1232) in Compton scattering off the
  %nucleon,''
  Phys.\ Rev.\  C {\bf 67}, 055202 (2003).
  % [nucl-th/0212024].
 % \href{http://arxiv.org/abs/nucl-th/0212024}{arXiv:nucl-th/0212024}.
  %%CITATION = PHRVA,C67,055202;%%

%\cite{Pascalutsa:2005vq}
\bibitem{Pascalutsa:2005vq} 
  V.~Pascalutsa and M.~Vanderhaeghen,
  %``Chiral effective-field theory in the Delta(1232) region: I. Pion electroproduction on the nucleon,''
  Phys.\ Rev.\ D {\bf 73}, 034003 (2006).
%  doi:10.1103/PhysRevD.73.034003
  %[hep-ph/0512244].
  %%CITATION = doi:10.1103/PhysRevD.73.034003;%%
  %88 citations counted in INSPIRE as of 26 Feb 2018
  
   %\cite{Christy:2007ve}
\bibitem{Christy:2007ve} 
  M.~E.~Christy and P.~E.~Bosted,
  %``Empirical fit to precision inclusive electron-proton cross- sections in the resonance region,''
  Phys.\ Rev.\ C {\bf 81}, 055213 (2010).
%  [arXiv:0712.3731 [hep-ph]].
  %%CITATION = ARXIV:0712.3731;%%

\bibitem{Holstein:1999uu} 
  B.~R.~Holstein, D.~Drechsel, B.~Pasquini, and M.~Vanderhaeghen,
  %``Higher order polarizabilities of the proton,''
  Phys.\ Rev.\ C {\bf 61}, 034316 (2000). 
  %[hep-ph/9910427].
  %%CITATION = HEP-PH/9910427;%%

\bibitem{Pasquini:2001yy} 
  B.~Pasquini, M.~Gorchtein, D.~Drechsel, A.~Metz, and M.~Vanderhaeghen,
  %``Dispersion relation formalism for virtual Compton scattering off the proton,''
  Eur.\ Phys.\ J.\ A {\bf 11}, 185 (2001). 
 % [hep-ph/0102335].
  %%CITATION = HEP-PH/0102335;%%  

\bibitem{Drechsel:2007if} 
  D.~Drechsel, S.~S.~Kamalov, and L.~Tiator,
  %``Unitary Isobar Model - MAID2007,''
  Eur.\ Phys.\ J.\ A {\bf 34}, 69 (2007).
%  [arXiv:0710.0306 [nucl-th]].
  %%CITATION = ARXIV:0710.0306;%%  

%\cite{Joo:2001tw}
\bibitem{Joo:2001tw} 
  K.~Joo {\it et al.} (CLAS Collaboration),
  %``Q**2 dependence of quadrupole strength in the gamma* p ---> Delta+(1232) ---> p pi0 transition,''
  Phys.\ Rev.\ Lett.\  {\bf 88}, 122001 (2002).
  %doi:10.1103/PhysRevLett.88.122001
%  [hep-ex/0110007].
  %%CITATION = doi:10.1103/PhysRevLett.88.122001;%%
  %210 citations counted in INSPIRE as of 24 Feb 2018

\bibitem{Gryniuk:2015eza} 
  O.~Gryniuk, F.~Hagelstein, and V.~Pascalutsa,
  %``Evaluation of the forward Compton scattering off protons: Spin-independent amplitude,''
  Phys.\ Rev.\ D {\bf 92}, 074031 (2015).
%  doi:10.1103/PhysRevD.92.074031
%  [arXiv:1508.07952 [nucl-th]].
  %%CITATION = doi:10.1103/PhysRevD.92.074031;%%

  \bibitem{Birse:2012eb} 
  M.~C.~Birse and J.~A.~McGovern,
  %``Proton polarisability contribution to the Lamb shift in muonic hydrogen at fourth order in chiral perturbation theory,''
  Eur.\ Phys.\ J.\ A {\bf 48}, 120 (2012).
%  [arXiv:1206.3030 [hep-ph]].
  %%CITATION = ARXIV:1206.3030;%%

\bibitem{Tomalak:2015hva} 
  O.~Tomalak and M.~Vanderhaeghen,
  %``Two-photon exchange correction to muon–proton elastic scattering at low momentum transfer,''
  Eur.\ Phys.\ J.\ C {\bf 76}, 125 (2016).
 % doi:10.1140/epjc/s10052-016-3966-3
 % [arXiv:1512.09113 [hep-ph]].
  %%CITATION = doi:10.1140/epjc/s10052-016-3966-3;%%  
   
 
   
\bibitem{Olive:2016xmw} 
  C.~Patrignani {\it et al.} (Particle Data Group Collaboration),
  %``Review of Particle Physics,''
  Chin.\ Phys.\ C {\bf 40}, 100001 (2016).
  %doi:10.1088/1674-1137/40/10/100001
  %%CITATION = doi:10.1088/1674-1137/40/10/100001;%%


    
\bibitem{McGovern:2012ew} 
  J.~A.~McGovern, D.~R.~Phillips, and H.~W.~Grie{\ss}hammer,
  %``Compton scattering from the proton in an effective field theory with explicit Delta degrees of freedom,''
  Eur.\ Phys.\ J.\ A {\bf 49}, 12 (2013).
%  [arXiv:1210.4104 [nucl-th]].
  %%CITATION = ARXIV:1210.4104;%%  


\bibitem{Lensky:2014efa}
  V.~Lensky and J.~A.~McGovern,
  %``Proton polarizabilities from Compton data using covariant chiral effective field theory,''
  Phys.\ Rev.\ C {\bf 89}, 032202 (2014). 
%  doi:10.1103/PhysRevC.89.032202
%  [arXiv:1401.3320 [nucl-th]].
  %%CITATION = doi:10.1103/PhysRevC.89.032202;%%
\bibitem{Alarcon:2013cba}
J.~M.~Alarc\'on, V.~Lensky, and V.~Pascalutsa,     
%{Chiral perturbation theory of muonic hydrogen Lamb shift: polarizability contribution
Eur.\ Phys.\ J.\ C {\bf 74}, 2852 (2014).
%%CITATION = ARXIV:1312.1219;%%"

\bibitem{Carlson:2011zd}
C.~E.~Carlson and M.~Vanderhaeghen,
%Higher order proton structure corrections to the Lamb shift in muonic hydrogen
Phys.\ Rev.\ A {\bf 84}, 020102 (2011).
%%CITATION = ARXIV:1101.5965;%%"

 \bibitem{Hagelstein2017}
 F.~Hagelstein, 
 %{\it Exciting Nucleons in Compton Scattering and Hydrogen-Like Atoms}, 
 Ph.D.\ thesis, Johannes Gutenberg-Universit{\"a}t, Mainz, 2017, arXiv:1710.00874 [nucl-th].

\bibitem{Pascalutsa:2007wz}
V.~Pascalutsa and M.~Vanderhaeghen, 
%Large-N(c) relations for the electromagnetic N to Delta(1232) transition}",
Phys.\ Rev.\ D {\bf 76}, 111501 (2007).
%      eprint         = "0711.0147",
%      archivePrefix  = "arXiv",
%      primaryClass   = "hep-ph",
%%CITATION = ARXIV:0711.0147;%%"

\bibitem{Bradford:2006yz}
R.~Bradford, A.~Bodek, H.~S.~Budd, and J.~Arrington,
%A New parameterization of the nucleon elastic form-factors
Nucl.\ Phys.\ Proc.\ Suppl.\ {\bf 159}, 127 (2006).
%"hep-ex/0602017",
%%CITATION = HEP-EX/0602017;%%"

%\cite{Antognini:2013jkc}
\bibitem{Antognini:2013jkc} 
  A.~Antognini, F.~Kottmann, F.~Biraben, P.~Indelicato, F.~Nez, and R.~Pohl,
  %``Theory of the 2S-2P Lamb shift and 2S hyperfine splitting in muonic hydrogen,''
  Ann.\ Phys.\  {\bf 331}, 127 (2013).
  %doi:10.1016/j.aop.2012.12.00
  %[arXiv:1208.2637 [physics.atom-ph]].
  %%CITATION = doi:10.1016/j.aop.2012.12.003;%%
  %81 citations counted in INSPIRE as of 06 Feb 2018

  \bibitem{Peset:2014jxa}
C.~Peset and A.~Pineda, Nucl.\ Phys.\ B {\bf 887}, 69 (2014).
%%CITATION = ARXIV:1406.4524;%%"

\bibitem{Lvov:1980wp} 
  A.~I.~L'vov,
  %``Compton Scattering on Proton at Energies Up to 400-{MeV} and Finite Energy Sum Rules,''
  Yad.\ Fiz.\  {\bf 34}, 1075 (1981)
  [Sov.\ J.\ Nucl.\ Phys.\  {\bf 34}, 597 (1981)].
  %%CITATION = SJNCA,34,597;%%

\bibitem{Lvov:1996rmi} 
  A.~I.~L'vov, V.~A.~Petrun'kin, and M.~Schumacher,
  %``Dispersion theory of proton Compton scattering in the first and second resonance regions,''
  Phys.\ Rev.\ C {\bf 55}, 359 (1997).
  %doi:10.1103/PhysRevC.55.359
  %%CITATION = doi:10.1103/PhysRevC.55.359;%%
  

  
  
  
\bibitem{Babusci:1998ww} 
  D.~Babusci, G.~Giordano, A.~I.~L'vov, G.~Matone, and A.~M.~Nathan,
%   ``Low-energy Compton scattering of polarized photons on polarized nucleons,''
  Phys.\ Rev.\ C {\bf 58}, 1013 (1998). 
  %[hep-ph/9803347].
  %%CITATION = HEP-PH/9803347;%%




\bibitem{Guichon:1995pu} 
  P.~A.~M.~Guichon, G.~Q.~Liu, and A.~W.~Thomas,
  %``Virtual Compton scattering and generalized polarizabilities of the proton,''
  Nucl.\ Phys.\ A {\bf 591}, 606 (1995).
 % [nucl-th/9605031].
  %%CITATION = NUCL-TH/9605031;%%
 
  
\bibitem{Drechsel:1998zm} 
  D.~Drechsel, G.~Kn{\"o}chlein, A.~Y.~Korchin, A.~Metz, and S.~Scherer,
  %``Low-energy and low momentum representation of the virtual Compton scattering amplitude,''
  Phys.\ Rev.\ C {\bf 58}, 1751 (1998). 
  %doi:10.1103/PhysRevC.58.1751
  %[nucl-th/9804078].
  %%CITATION = doi:10.1103/PhysRevC.58.1751;%%


  
\end{thebibliography}
\end{document}